\documentclass[10.5pt,letterpaper]{article}

\usepackage[utf8]{inputenc} 
\usepackage{eurosym}
\usepackage[svgnames]{xcolor}
\usepackage{amsmath,amssymb,amsfonts,amsthm}
\usepackage{mathtools,old-arrows,verbatim,mathrsfs,cite}

\usepackage{graphicx}
\graphicspath{{fig/}} 

\usepackage{epstopdf,epsfig,changepage}

\usepackage{bm}
\usepackage[top=2.5cm, bottom=3cm, left=3.5cm, right=3.5cm,
           heightrounded,marginparwidth=1.5cm, marginparsep=1cm]{geometry}
\usepackage[shortlabels]{enumitem}
\usepackage[square,numbers,merge,comma,sort&compress]{natbib} 
\makeatletter
\def\NAT@spacechar{\,}  
\makeatother
\usepackage{comment}
\usepackage{relsize,setspace,moresize}
\usepackage{latexsym,mathrsfs,calligra,aurical}
\usepackage{float,appendix,xargs,extarrows,empheq,url}

\usepackage[labelsep=colon,skip=5pt,aboveskip=5pt,belowskip=10pt]{subcaption}

\usepackage[svgnames]{xcolor}

\usepackage{jheppub}
\hypersetup{
    bookmarks=true,pdfmenubar=true,pdffitwindow=false,pdfpagemode={UseNone},
    pdfstartview={FitH},colorlinks=true,bookmarks=true,bookmarksnumbered=true,
    plainpages,linktoc=page,citecolor=blue,filecolor=black,
    linkcolor=Crimson,urlcolor=Green}
\usepackage{footnotebackref}
\usepackage{hypernat}

\DeclareFixedFont\trfont{OT1}{phv}{b}{sc}{11}
\def\={\:=\:}
\newcommand{\N}{\mathcal{N}}
\newcommand\pa{\partial}

\newcommand\nv{n_\text{v}}
\newcommand\ns{n_\text{s}}
\newcommand\Ms{\mathscr{M}}
\newcommand\zb{{\bar{z}}}
\newcommand\ib{{\bar{\imath}}}
\newcommand\jb{{\bar{\jmath}}}
\newcommand\Mscal{\mathscr{M}_\text{scal}}

\renewcommand{\[}{\left[}

\newcommandx{\tcr}[1]{\textcolor{Crimson}{#1}}
\newcommandx{\ETh}[2][1=M,2=\alpha,usedefault]{\Theta_{#1}{}^{#2}}
\newcommandx{\overbar}[1]{\mkern1.5mu\overline{\mkern-2.0mu#1\mkern-2.0mu}\mkern1.5mu}
\newcommandx{\overbarM}[1]{\mkern6.0mu\overline{\mkern-5.5mu#1\mkern-3.5mu}\mkern1.5mu}
\newcommandx{\overbarcal}[1]{\mkern6.0mu\overline{\mkern-5.5mu#1\mkern-1.0mu}\mkern1.5mu}
\DeclareFixedFont\trfont{OT1}{phv}{b}{sc}{11}
\DeclareMathAlphabet{\mathpzc}{OT1}{pzc}{m}{it}
\DeclareMathAlphabet{\mathcal}{OMS}{cmsy}{m}{n}
\DeclareSymbolFontAlphabet{\Scr}{rsfs}
\DeclareMathAlphabet{\mathbold}{U}{BOONDOX-ds}{m}{n}
\SetMathAlphabet{\mathbold}{bold}{U}{BOONDOX-ds}{b}{n}
\DeclareMathAlphabet{\mathcalboondox}{U}{BOONDOX-calo}{m}{n}
\SetMathAlphabet{\mathcalboondox}{bold}{U}{BOONDOX-calo}{b}{n}
\DeclareMathAlphabet{\mathbcalboondox}{U}{BOONDOX-calo}{b}{n}

\title{\centering\boldmath\LARGE\bfseries{%
        Exact holographic RG flows in extended SUGRA%
        }\vspace{1.25em}}

\author[a]{A.\ Anabalón,}
\emailAdd{andres.anabalon@uai.cl}

\author[b]{D.\ Astefanesei,}
\emailAdd{dumitru.astefanesei@pucv.cl}

\author[a]{D.\ Choque,}
\emailAdd{david.choque@pucv.cl}

\author[c,d]{A.\ Gallerati}
\emailAdd{antonio.gallerati@polito.it}

\author[c,d]{and M.\ Trigiante}
\emailAdd{mario.trigiante@polito.it}

\bigskip

\affiliation[a]{%
\makebox[\textwidth][l]%
{Universidad Adolfo Ibàñez, Dep.\ de Ciencias, Fac.\ Artes Liberales. Av.\ Padre Hurtado 750, Viña del Mar, Chile}%
}

\affiliation[b]{Pontificia Universidad Cat\'{o}lica de Valpara\'{i}so,
Instituto de F\'{i}sica, Av.\ Brasil 2950, Valpara\'{i}so, Chile}

\affiliation[c]{Politecnico di Torino, Dipartimento DISAT. Corso Duca degli Abruzzi 24, 10129 Torino, Italy}

\affiliation[d]{Istituto Nazionale di Fisica Nucleare (INFN), sez.\ TO. Via Pietro Giuria 1, Torino, Italy \bigskip}

\bigskip

\abstract{%
We present a family of exact planar hairy neutral black hole solutions in extended supergravity with Fayet-Iliopoulos (FI) terms. We consider a model where the magnetic part of FI sector vanishes and obtain the superpotential at finite temperature in analytic form. Then, we discuss the thermodynamics and some holographic properties of these solutions. We regularize the action by two different methods, one with gravitational and scalar counterterms and
the other using the thermal superpotential as a counterterm, and compute the holographic stress tensor. We also construct the $c$-function of the
corresponding RG flow and obtain an exact holographic $\beta$-function for this model.}

\date{}

\begin{document}

\maketitle


\section{Introduction}\label{sec:Intro}
The AdS-CFT correspondence \cite{Maldacena:1997re} provides a striking
geometric picture for quantum field theories with a gravity dual. Usually
referred to as `holographic duality', this proposal explicitly relates a
non-gravitational field theory with a (super)gravity theory with one extra
spatial dimension. Using this correspondence, one can gain new insights into
both of the theories on either side of the duality.

It is often stated that the dual field theory `lives on the boundary' that,
in fact, means  that the theory is at the UV critical point. More generally,
there exists a UV/IR relation that identifies (super)gravity degrees of
freedom at large (small) radius to those in the field theory at high (low)
energy. The renormalization group (RG) is a powerful method for constructing relations between theories at different length scales. Since the energy scale of the `boundary theory' corresponds to the
radial direction in the bulk spacetime \cite{Susskind:1998dq,Peet:1998wn},
the geometric radial flow can be holographically interpreted as the
renormalization group flow of the dual field theory \cite{deBoer:1999tgo,deBoer:2000cz}. Concretely, this fundamental feature of
AdS-CFT duality is based on the fact that the boundary values of bulk fields
determine the dual field theory couplings, combined with the fact that
the motion in the radial direction is related to scaling in the dual field
theory.

The original proposal of Maldacena stated that the (most) supersymmetric $\mathcal{N}=4$ four-dimensional $\mathrm{SU}(\N)$ gauge theory is equivalent to Type IIB superstring theory on its maximally supersymmetric background of the form $\text{AdS}_5\times {\rm S}^5$. The massless sector of this superstring theory, compactified on the five-sphere ${\rm S}^5$, is described by an effective five-dimensional supergravity, which is a maximally supersymmetric one with gauge group ${\rm SO}(6)$. This five-dimensional supergravity theory, in light of the conjectured duality, captures a suitable sector of the dual SYM theory in one dimension less. This gauge theory is special because its $\beta$-function vanishes, resulting then in a conformal invariant (quantum) field theory. In the framework of string theory, where the AdS-CFT duality is best understood, simple backgrounds are both supersymmetric and conformally invariant. It is interesting, though, to
find new gravity solutions that could describe non-trivial RG flows. For
example, it is well known that by adding certain relevant perturbations the
theory may flow from the UV fixed point CFT to a new fixed point (a different
conformal field theory) in the infrared \cite{Distler:1998gb,Girardello:1998pd,Freedman:1999gp,Girardello:1999bd}. Generally, one way to obtain a non-trivial $\beta$-function is to consider a model with a dilaton being non-constant. While domain walls connecting two different AdS vacua are by now routinely found, exact non-singular flows at finite temperature can not be constructed easily.\par\smallskip
In this paper, we present exact neutral hairy black hole (BH) solutions in $D=4$,  \,$\mathcal{N}=2$ gauged supergravity with electric FI terms \cite{Anabalon:2017yhv}.%
\footnote{Various neutral \cite{Henneaux:2002wm,Martinez:2004nb,Hertog:2004dr,Anabalon:2012ta,Acena:2012mr, Acena:2013jya,Wen:2015xea,Fan:2015tua,Kichakova:2015nni} and charged \cite{Anabalon:2013sra,Anabalon:2013eaa,Lu:2013ura,Astefanesei:2018vga,Anabalon:2020pez} static black hole configurations with scalar hair have been considered in the literature.}
For some consistent supergravity truncations with exact hairy black hole solutions (see for instance \cite{Feng:2013tza,Lu:2014fpa,Faedo:2015jqa,Anabalon:2019tcy}), the related scalar potentials are particular cases of the general one obtained in \cite{Anabalon:2017yhv}. The specific form of the dilaton potential\footnote{When the cosmological constant is turned off, the potential is still non-trivial and charged hairy black holes, which are asymptotically flat, exist when considering this potential \cite{Anabalon:2013qua} in Einstein-Maxwell-dilaton theory. Interestingly, it was proven in \cite{Astefanesei:2019mds,Astefanesei:2019qsg,Astefanesei:2020xvn} that this particular self-interaction for the scalar field is essential for the thermodynamic and dynamic stability of the asymptotically flat hairy black hole. } in this model is connected with the existence of exact solutions and, also, with a black hole solution generating technique that is going to be very useful for our work. The scalar
field has mixed boundary conditions that preserve the isometries of AdS
spacetime. In the context of AdS-CFT duality, these generalized boundary
conditions have a nice interpretation as multi-trace deformations of the
boundary CFT \cite{Witten:2001ua}. For our solutions, the mixed boundary conditions of the scalar field correspond to adding a triple-trace operator to the dual field theory action. We are then going to focus on thermal holographic properties of the
corresponding RG flow. For the radial flow, the monotonic $c$-function can be related to a thermal superpotential that allows one to  recast  the  second  order  bulk  gravity  equations  into first order equations.

The paper is structured as follows.  In Section \ref{sec:Sugra}, we describe the supergravity framework and present a consistent truncation for the dilaton. In Section \ref{sec:HairyBH}, we present the exact hairy BH solutions and obtain their thermodynamic properties using holographic techniques. In Section \ref{sec:Holography},  after a brief review of the construction of the RG flow at finite temperature, we obtain the thermal superpotential and present some concrete applications within AdS-CFT duality. Finally, Section \ref{sec:Discussion} contains a discussion of our results with the emphasis on physical interpretations.


\section{Gauged supergravity framework} \label{sec:Sugra}
The construction of stationary black hole configurations is motivated by the study of classical general relativity solutions as well as AdS-CFT duality. These studies suggest that conditions for the existence of hairy black hole solutions comprise suitable scalar field self-interaction properties, encoded in a scalar potential, together with an appropriate gravitational interaction determining the near-horizon behaviour as well as the far-region hair physics. This implies a probable connection between the integrability of the equations of motion and the explicit form of the scalar potential. An embedding of the scalar potential itself in a supergravity model is important, since many physical aspects of the theory can be better understood. In this section we describe in great detail the supergravity theory we are going to use to obtain exact hairy black hole solutions.

\paragraph{Scalar potential.}
In a supergravity theory, supersymmetry constrains the form of the scalar potential, allowing certain classes of solutions to be described by first order  `gradient flow' equations, easier to handle. In particular, critical points of the scalar potential in $D=4$ supergravity define the asymptotic features of the black hole solution at radial infinity and the dual CFT.

A scalar potential together with fermion mass terms can be introduced in a supergravity theory without manifestly breaking supersymmetry only under certain conditions \cite{Nilles:1983ge,Wess:1992cp}.
In extended supergravities, the only known mechanism for introducing a non-trivial scalar potential without explicitly breaking supersymmetry is the so-called \emph{gauging procedure} \cite{deWit:1981sst,deWit:1982bul,Hull:1984vg,Hull:1984qz}. The latter can be seen as a deformation of an ungauged theory, with the same amount of supersymmetry and field content, where a suitable subgroup of the global symmetry group of the Lagrangian is promoted to \emph{local} symmetry, to be gauged by the vector fields.
The original (ungauged) Lagrangian is then modified, replacing the abelian vector field strengths by non-abelian ones, introducing proper covariant derivatives, Yukawa terms and a suitable scalar potential. The coupling of the (formerly abelian) vector fields to the new local gauge group provides matter fields that are charged under the new local gauge symmetry. Theories featuring an internal gauge symmetry and, related to it, a non vanishing scalar potential, are generically referred to as \emph{gauged supergravities}.

\paragraph{Embedding tensor.}
The above mentioned gauging procedure will in general break the global symmetry group of the original ungauged theory: this global symmetry, acting as a generalized electric-magnetic duality, is broken by the introduced minimal couplings, which only involve the electric vector fields.
As a consequence of this, in a gauged model we loose track of the string/M-theory dualities, which are conjectured to be encoded in the global symmetries of the ungauged theory \cite{Hull:1994ys}.\par
The above issue can be avoided using the \emph{embedding tensor formulation} of the gauging procedure \cite{Cordaro:1998tx,Nicolai:2000sc,deWit:2002vt,deWit:2004nw,deWit:2005hv,deWit:2005ub,deWit:2007kvg,deWit:2008ta,Samtleben:2008pe,Trigiante:2016mnt,Gallerati:2016oyo,Gallerati:2019mzs}, in which all deformations involved are encoded in a single object, the embedding tensor $\Theta$, which is itself covariant with respect to the global symmetries of the ungauged model. This procedure allows to formally restore the symmetries at the level of the (gauged) field equations and Bianchi identities, provided the embedding tensor is transformed together with the other fields. However, since the embedding tensor is a non-dynamical object, whose entries can be regarded as background quantities, a transformation on it will map a model into a different one. Therefore, in the embedding tensor formulation of gauged supergravities, global symmetries of the ungauged theory now act as equivalences between different gauged models.
%

\paragraph{Fayet-Iliopoulos terms.}
In supergravity theories, the scalar fields in the Lagrangian are typically described by a non-linear sigma-model, that is, they are coordinates of a non-compact, Riemannian $\ns$--dimensional differentiable manifold, the target
space $\Mscal$. We shall restrict ourselves to the case in which the latter is a homogeneous, symmetric manifold of the form%
\footnote{%
all scalar manifolds of $\mathcal{N}>2$ supergravity theories are of this kind, while $\mathcal{N}=2$ can have homogeneous non-symmetric and even non-homogeneous scalar manifolds
}
\begin{equation}
\Mscal\:\sim\:G/H\;,
\end{equation}
\sloppy
where $G$ is the manifold isometry group and $H$ is the isotropy group of the origin $\mathcal{O}$. 
The scalar manifolds $\Mscal$ spanned by the scalar fields in the vector multiplets of $\mathcal{N}=2$ theories, as well as the scalar manifolds in all $\mathcal{N}>2$ four-dimensional theories, are endowed with a \emph{flat symplectic bundle}. As a consequence of this, with each point of these spaces $\Mscal$ a characteristic symmetric symplectic matrix $\mathcal{M}_{MN}$ is defined, determining a metric on the symplectic fiber and encoding all information about the non-minimal coupling between the scalar fields and the vectors. Moreover, within the flat symplectic structure, each isometry of the manifold is naturally associated with a constant symplectic matrix, with respect to which $\mathcal{M}_{MN}$ transforms as a metric under the action of the isometry.\par
We shall be interested in extended $\N=2$ theories, in which scalar fields may sit either in the vector multiplets or in the hypermultiplets (that are part of the fermionic sector of the theory). The former scalars span a \emph{special K\"ahler manifold} $\Ms_\textsc{sk}$, while the latter, named hyper-scalars, parameterize a quaternionic K\"ahler one $\Ms_\textsc{qk}$ \cite{Strominger:1990pd,Bagger:1983tt,Lauria:2020rhc}.
The scalar manifold is always factorized in the product of the two, while the isotropy group $H$ of the scalar manifold splits according to $H=H_\text{R}\times H_\text{matt}$, where $H_\text{R}=\mathrm{U}(2)$ is the R-symmetry group and $H_\text{matt}$ acts on the matter fields in the vector and hypermultiplets.\par
In the absence of hypermultiplets, the $\mathrm{SU}(2)$ part of the R-symmetry group $H_\text{R}$ becomes a global symmetry of the theory which can still be gauged, the gauging of this symmetry being described by a (constant) embedding tensor whose components are known as \emph{Fayet-Iliopoulos terms} (FI terms).
If the special K\"ahler isometries are not involved in the gauging, the constraints imply that only a $\mathrm{U}(1)$ subgroup of $\mathrm{SU}(2)$ can be gauged. In this case, the embedding tensor has only one non-vanishing component and the resulting theory is deformed by the introduction of abelian electric-magnetic FI terms defined by a constant symplectic vector $\theta_M$, which encodes all the gauge parameters%
\footnote{%
even if we introduce both electric and magnetic gaugings to maintain duality
covariance, the duality group will always allow us to reduce to the case with only electric gaugings turned on
}.\par\smallskip%
In the following we will consider a class of $\N=2$ supergravities coupled to a single vector multiplet in the presence of FI terms. In particular, we will analyse a consistent dilaton truncation of the model with an explicit form for the scalar potential, discussing also how to explicitly express the latter in terms of the chosen FI quantities.
The formulation will lead to an asymptotically AdS, regular hairy black hole class of solutions. This kind of models can also feature unexpected symmetries involving parameter transformations with non-trivial action, providing a new solution generating technique in asymptotically AdS spacetimes%
\footnote{%
this is not a generic assumption, since in asymptotically AdS black holes the solutions generating technique \cite{Breitenlohner:1987dg,Cvetic:1995kv,Cvetic:1996kv,Gaiotto:2007ag,Bergshoeff:2008be,Bossard:2009at,Fre:2011uy,Andrianopoli:2013kya,Andrianopoli:2013jra,Andrianopoli:2013ksa,Gallerati:2019mzs}, based on the global symmetry group of the ungauged theory, can no longer be applied in a gauged theory, due to the non-trivial duality action on the embedding tensor \cite{Gallerati:2016oyo,Trigiante:2016mnt}%
} \cite{Anabalon:2017yhv}.

\subsection{Gauged supergravity with FI terms}
Let us consider an extended \,$\mathcal{N}=2$ supergravity theory in four dimensions, coupled to $\nv$ vector multiplets and no hypermultiplets, in the presence of Fayet--Iliopoulos (FI) terms. The model describes $\nv$ vector fields $A^\Lambda_\mu$,\, ($\Lambda=1,\dots,\nv$) and $\ns=\nv-1$ complex scalar fields $z^i$ ($i=1,\dots,\ns$)%
\footnote{%
in our previous work \cite{Anabalon:2017yhv} $\nv$ denoted the number of vector multiplets of the theory, while, in the more general formulation of the present work, it directly provides the total number of vector fields
}.
The bosonic gauged Lagrangian is written as
\begin{equation}
\frac{1}{e}\,\mathscr{L}_{\textsc{bos}}~=\,
-\frac{R}{2}
\:+\:g_{i\bar{\jmath}}\,\partial_\mu z^i\,\partial^\mu \zb^{\bar{\jmath}}
\:+\:\frac{1}{4}\,\mathcal{I}_{\Lambda\Sigma}(z,\zb)\,F^\Lambda_{\mu\nu}\,F^{\Sigma\,\mu\nu}
\:+\:\frac{1}{8\,e}\,\mathcal{R}_{\Lambda\Sigma}(z,\zb)\,\varepsilon^{\mu\nu\rho\sigma}\,F^\Lambda_{\mu\nu}\,F^{\Sigma}_{\rho\sigma}
\:-\:V(z,\zb)\;,
\label{eq:boslagr}
\end{equation}
with the $\nv$ vector field strengths:
\begin{equation}
F^\Lambda_{\mu\nu}\=\partial_\mu A^\Lambda_\nu-\partial_\nu A^\Lambda_\mu\;.
\end{equation}
The $\ns$ complex scalars $z^i$ couple to the vector fields through the real symmetric matrices $\mathcal{I}_{\Lambda\Sigma}(z,\zb)$, $\mathcal{R}_{\Lambda\Sigma}(z,\zb)$ (non-minimal couplings) and
span a special K\"ahler manifold $\mathscr{M}_\textsc{sk}$, the scalar potential $V(z,\zb)$ originating from electric-magnetic FI terms.
The presence of $V(z,\zb)$ amounts to gauging a $\mathrm{U}(1)$-symmetry of the corresponding ungauged model (with no FI terms), implying minimal couplings of the vectors to the fermion fields only.

\subsubsection{Special geometry.}\label{subsubsec:geometry}
A special K\"ahler manifold $\mathscr{M}_\textsc{sk}$ is the class of target spaces that are spanned by the complex scalar fields sitting in the vector multiplets of an $\mathcal{N}=2$ four-dimensional supergravity.\par
The geometrical properties of $\mathscr{M}_\textsc{sk}$ are described in terms of a \emph{holomorphic section} $\Omega^M(z^i)$ of the characteristic bundle defined over it. The latter is expressed by the product of a symplectic-bundle and a holomorphic line-bundle. The components of the section are written as
\begin{equation}
\Omega^M=
\left(\begin{matrix}
\mathcal{X}^\Lambda \cr \mathcalboondox{F}_\Lambda
\end{matrix}\right)\;, \qquad\qquad
\Lambda=1,\,\dots,\nv\;,
\end{equation}
while the {K\"ahler potential} and the {K\"ahler metric} have the following general form
\begin{equation} \label{eq:Kahlmetr}
\begin{split}
\mathcal{K}(z,\zb)&~=\,
    -\log\left[\,i\;\overbar{\Omega}^T\,\mathbb{C}\;\Omega\,\right]~=\,
    -\log\left[\,i\,\left(\overbarcal{\mathcal{X}}^\Lambda\,\mathcalboondox{F}_\Lambda-{\mathcal{X}}^\Lambda\,\overbar{\mathcalboondox{F}}_\Lambda\,\right)\right]\;,
\\[1.5\jot]
g_{i\bar{\jmath}}~&=~\partial_i \partial_{\bar{\jmath}}\mathcal{K}\;.
\end{split}
\end{equation}
The choice of $\Omega^M(z^i)$ can be used to fix the symplectic frame (basis of the symplectic fiber space) and, consequently, the non-minimal couplings of the scalars to the vector field strengths in the Lagrangian. In the special coordinate frame, the lower entries $\mathcalboondox{F}_\Lambda$ of the section can be expressed as the gradient, with respect to the upper components $\mathcal{X}^\Lambda$, of a characteristic \emph{prepotential function} $\mathcal{F}(\mathcal{X}^\Lambda)$:
\begin{equation}
\mathcalboondox{F}_\Lambda\=\frac{\partial\mathcal{F}}{\partial \mathcal{X}^\Lambda}\;.
\end{equation}
The above function $\mathcal{F}(\mathcal{X}^\Lambda)$ is required to be homogeneous of degree two. The upper entries $\mathcal{X}^\Lambda(z^i)$ are defined modulo multiplication times a holomorphic function and, in this frame, can be used as projective coordinates to describe the manifold; this means that, in a local patch in which
$\mathcal{X}^0\neq 0$, we can identify the scalar fields with the ratios $z^i=\mathcal{X}^i/\mathcal{X}^0$.\par\smallskip
A field $\Phi(z,\zb)$ on the K\"ahler manifold is expressed as a section of a $\mathrm{U}(1)$-bundle of weight $p$ if it transforms under a K\"ahler transformation as
\begin{equation}
\Phi(z,\zb)\;\;\rightarrow\;\;e^{i\,p\,\operatorname{Im}[f]}\,\Phi(z,\zb)\;.
\end{equation}
We can define an associated $\mathrm{U}(1)$-\emph{covariant derivative} on the bundle as
\begin{equation}
\begin{split}
\mathcal{D}_i \Phi \,\equiv\,\left(\partial_i+\frac{p}{2}\,\partial_i\mathcal{K}\right)\Phi\;,
\qquad\quad
\mathcal{D}_{\ib}\Phi\,\equiv\,\left(\partial_\ib-\frac{p}{2}\,\partial_\ib\mathcal{K}\right)\Phi\;.
\end{split}
\end{equation}
and define a \emph{covariantly holomorphic vector} $\mathcal{V}^M$
\begin{equation}
\mathcal{V}^M\=e^{\frac{\mathcal{K}}{2}}\,\Omega^M\=
    \left(\begin{matrix}
          L^\Lambda \cr M_\Lambda
          \end{matrix}\right)\;,
\label{eq:VM}
\end{equation}
which is section of the $\mathrm{U}(1)$-line bundle with weight $p=1$, satisfying the property:
\begin{equation}
\mathcal{D}_{\ib}\,\mathcal{V}^M\=
    \left(\partial_{\bar{\imath}}-\frac{1}{2}\,\partial_{\bar{\imath}}\mathcal{K}\right)\mathcal{V}^M\=0\:.
\end{equation}
We can also introduce the quantities
\begin{equation}
\mathcal{U}_i^M\:\equiv\:\mathcal{D}_i\,\mathcal{V}^M \=
     \left(\partial_i+\frac{1}{2}\,\partial_i\mathcal{K}\right)\,\mathcal{V}^M\=
     \left(\begin{matrix}
            f_i^\Lambda  \cr  h_{i\Lambda}
           \end{matrix}\right)\;,
\label{eq:Ui}
\end{equation}
where $\mathcal{D}_{i},\,\mathcal{D}_{\ib}$ are the above $\mathrm{U}(1)$-covariant derivatives.
The scalar potential $V(z,\zb)$, expressed in terms of the new quantities, reads:
\begin{equation}
V\=\left(g^{i\bar{\jmath}}\,\mathcal{U}_i^M\,\overbar{\mathcal{U}}_{\bar{\jmath}}^N
         -3\,\mathcal{V}^M\,\overbar{\mathcal{V}}^N\right)\theta_M\,\theta_N\,=\,
    -\frac{1}{2}\,\theta_M\,\mathcal{M}^{MN}\,\theta_N-4\,\mathcal{V}^M\,\overbar{\mathcal{V}}^N\theta_M\,\theta_N\;,
\label{eq:VpotUV}
\end{equation}
where $\mathcal{M}^{MN}$, and its inverse $\mathcal{M}_{MN}$, are symplectic, symmetric, negative definite matrices encoding the non-minimal couplings of the scalar fields $z^i$ to the vectors. In particular $\mathcal{M}_{MN}$ is expressed as:
\begin{equation}
\mathcal{M}_{MN}\:\equiv\:
\left(
\begin{matrix}
(\mathcal{R}\mathcal{I}^{-1}\mathcal{R}+\mathcal{I})_{\Lambda\Sigma} &\;\;-(\mathcal{R}\mathcal{I}^{-1})_\Lambda{}^\Gamma \\ -(\mathcal{I}^{-1}\mathcal{R})^\Delta{}_\Sigma & (\mathcal{I}^{-1})^{\Delta\Gamma} \\
\end{matrix}
\right)\;,
\label{eq:Mphi}
\end{equation}
and the matrices $\mathcal{I},\,\mathcal{R}$ are those involved in the vector field strengths terms in \eqref{eq:boslagr}. The potential \eqref{eq:VpotUV} can be obtained in terms of a \emph{complex superpotential}
\begin{equation}
\mathcal{W}\=\mathcal{V}^M\,\theta_M\;,
\label{eq:W}
\end{equation}
section of the $\mathrm{U}(1)$-bundle with $p=1$, as:
\begin{equation}
V\=g^{i\jb}\,\mathcal{D}_i\mathcal{W}\;\mathcal{D}_{\jb}\overbar{\mathcal{W}}-3\,|\mathcal{W}|^2\;.
\label{eq:VW}
\end{equation}
It is also possible to define a \emph{real superpotential} $\mathcalboondox{W}=|\mathcal{W}|$ in terms of which the potential reads:
\begin{equation}
V\=4\,g^{i\bar{\jmath}}\,\partial_i\mathcalboondox{W}\,\partial_{\bar{\jmath}}\mathcalboondox{W}
    -3\,\mathcalboondox{W}^2\;.
\label{eq:VWre}
\end{equation}
The introduced $\theta_M$ terms transform in a symplectic representation of the isometry group $G_\textsc{sk}$ of $\mathscr{M}_\textsc{sk}$ on contravariant vectors. These Fayet-Iliopulos terms are the analogs of electric and magnetic charges; however, the latter can be considered as solitonic charges of the solution, while the former are background quantities actually entering the Lagrangian. Moreover the FI terms do not define vector-scalar minimal couplings but only fermion-vector ones.

\subsection{The model}
Let us focus on an $\mathcal{N}=2$ theory with no hypermultiplets and a single vector multiplet ($\nv=1$) with a complex scalar field $z$.
The geometry of the special K\"ahler manifold is described in terms of a prepotential of the form:
\begin{equation}
\mathcal{F}(\mathcal{X}^\Lambda)~=-\frac{i}{4}\:\left(\mathcal{X}^0\right)^{n}\left(\mathcal{X}^1\right)^{2-n}\;.
\end{equation}
the coordinate $z$ being identified with the ratio $\mathcal{X}^1/\mathcal{X}^0$. For special values of $n$, the model turns out to be a consistent truncation of the STU model. The latter is an $\mathcal{N}=2$ supergravity coupled to $\nv=3$ vector multiplets and is described, in a suitable symplectic frame, by the prepotential function:
\begin{equation}
\mathcal{F}_\textsc{stu}(\mathcal{X}^\Lambda)~=-\frac{i}{4}\,\sqrt{\mathcal{X}^0\,\mathcal{X}^1\,\mathcal{X}^2\,\mathcal{X}^3}\;,
\end{equation}
with a symmetric scalar manifold of the form $\mathscr{M}_\textsc{stu}=\big(\mathrm{SL}(2,\mathbb{R})/\mathrm{SO}(2)\big)^3$, spanned by three complex scalars $z^i=\mathcal{X}^i/\mathcal{X}^0$, $i=1,2,3$. This model is, in turn, a consistent truncation of the maximal $\mathcal{N}=8$ theory in $D=4$.\par
For the special value $n=1/2$, our model corresponds to the $z^3$--model, whose manifold is $\mathrm{SL}(2,\mathbb{R})/\mathrm{SO}(2)$ and is embedded in that of the STU model through the identification $z^1=z^2=z^3=z$.\;
If we set $\mathcal{X}^0=1$, the holomorphic section $\Omega^M$ of the theory under consideration reads:
\begin{equation}
\Omega^M=
\left(\begin{matrix}
1  \cr  z  \cr -\dfrac{i}{4}\,n\,z^{2-n} \cr -\dfrac{i}{4}\,(2-n)\,z^{1-n}
\end{matrix}\right)\;,
\end{equation}
and the K\"ahler potential $\mathcal{K}$ has the expression
\begin{equation}
e^{-\mathcal{K}}\=\frac{1}{4}\,z^{1-n}\,\big(n\,z-(n-2)\,\bar{z}\big)\:+\:\text{c.c.}
\end{equation}
The theory is then deformed by the introduction of abelian electric-magnetic
FI terms, defined by a constant symplectic vector $\theta_M=\left(\theta_1,\,\theta_2,\,\theta_3,\,\theta_4\right)$, encoding the gauge parameters of the model. Having found the explicit expressions for the section $\Omega^M$ and the K\"ahler potential $\mathcal{K}$, the scalar potential $V(z,\zb)$ can be read from \eqref{eq:VpotUV}, using \eqref{eq:VM} and \eqref{eq:Ui}.\par\smallskip
If we express the scalar $z$ in terms of a dilaton field $\varphi$ and an axionic field $\chi$
\begin{equation}
z=e^{\lambda\,\varphi}+i\,\chi\,,
\end{equation}
the truncation $\chi=0$ to the dilaton field $\varphi$ is consistent provided
\begin{equation}
(2-n)\,\theta_1\,\theta_3 - n\,\theta_2\, \theta_4\=0\,,
\label{eq:construnc}
\end{equation}
and the metric restricted to the dilaton reads:
\begin{equation}
ds^2\=2\,g_{z\bar{z}}\;dz\,d\bar{z}\,\big\vert_{{\!\!}_{\!\chi=0\atop d\chi=0}}
=~\frac{1}{2}\lambda^2\,n\,(2-n)\,d\varphi^2\;,
\label{eq:modulimetric}
\end{equation}
and is positive provided $0<n<2$. If we then set
\begin{equation}
\lambda\=\sqrt{\frac{2}{n\,(2-n)}}\;,
\end{equation}
the kinetic term for $\varphi$ is canonically normalized.
The scalar potential has now the explicit form:
\begin{equation}
\begin{split}
V\left(\varphi\right)\:=\,&-2\,e^{\lambda\,\varphi\,(n-2)}\,\left(\frac{2\,n-1}{n}\,\theta_{1}^2
+4\,\theta_{1}\,\theta_{2}\;e^{\lambda\,\varphi}
+\frac{2\,n-3}{n-2}\,\theta_{2}^2\;e^{2\,\lambda\,\varphi}\right)-\\
&-\frac{1}{8}\;e^{-\lambda\,\varphi\,(n-2)}\,\Big(\left(2\,n-1\right)\,n\,\theta_{3}^{2}
-4\,\theta_{3}\,\theta_{4}\,n\,\left(n-2\right)\,e^{-\lambda\,\varphi}
+\left(n-2\right)\,\left(2\,n-3\right)\,\theta_{4}^{2}\;e^{-2\,\lambda\,\varphi}\Big)\,,
\label{eq:Vpotdil}
\end{split}
\end{equation}
as a function of the dilaton only.\par

Let us remark that, in general, the truncation to the dilaton ($\chi=0$) is consistent at the level of scalar potential, but not of superpotential. In fact, if we consider the real superpotential $\mathcalboondox{W}=\lvert\mathcal{W}\rvert$, we find that in general
\begin{equation}
\partial_\chi\mathcalboondox{W}\big\rvert_{{}_{\chi=0}}\,\neq\,0\;,
\label{eq:nonconsistentWre}
\end{equation}
that, in turn, implies that the dilaton truncation cannot be extended at the level of the real superpotential. This means that, the scalar potential $V(\varphi)$ in \eqref{eq:Vpotdil} cannot be expressed in terms of $\mathcalboondox{W}\rvert_{{}_{\chi=0}}$ and its derivative with respect to $\varphi$, as it can ascertained by restricting \eqref{eq:VWre} to $\chi=0$, since on the r.h.s.\ a term depending on $\partial_\chi\mathcalboondox{W}\rvert_{{}_{\chi=0}}$ would appear in the expression of $V(\varphi)$. As we shall show below, this is not the case when $\alpha=0$, see Subsect.\ \ref{subsubsec:alpha0}.

\paragraph{Symmetries.}
The potential is invariant under the simultaneous transformations
\begin{equation}
z\rightarrow \frac{1}{z}\;,\quad\; \theta_{1}\rightarrow\pm\,\frac{n}{4}\,\theta_{3}\;,\quad\;
\theta_{2}\rightarrow\pm\,\frac{2-n}{4}\,\theta_{4}\;,\quad\;
\theta_{3}\rightarrow\mp\,\frac{4}{n}\,\theta_{1}\;,\quad\;
\theta_{4}\rightarrow\mp\,\frac{4}{2-n}\,\theta_{2}\;,
\end{equation}
implying the transformation $\varphi\rightarrow -\varphi$ in the dilaton truncation.
%
%
The potential is also invariant under
\begin{equation}
\theta_{1}\rightarrow\pm\,\frac{n}{4}\,\theta_{4}\;,\quad\;
\theta_{2}\rightarrow\pm\,\frac{2-n}{4}\,\theta_{3}\;,\quad\;
\theta_{3}\rightarrow\mp\,\frac{4}{n}\,\theta_{2}\;,\quad\;
\theta_{4}\rightarrow\mp\,\frac{4}{2-n}\,\theta_{1}\;,\quad\;
n \rightarrow 2-n\;.
\end{equation}

\medskip

\subsubsection{Simplifying the potential}
We now perform the shift
\begin{equation}
\varphi\,\rightarrow\,\varphi-\frac{2\,\nu}{\lambda\,(\nu+1)}\,\log(\theta_2\,\xi)\,,
\label{eq:phishift}
\end{equation}
and redefine the FI terms as:
\begin{equation}
\theta_{1}=\frac{\nu+1}{\nu-1}\;\theta_{2}^{-\frac{\nu-1}{\nu+1}}\,\xi^{-\frac{2\nu}{\nu+1}}\,,\qquad
\theta_{3}=2\,\alpha\left(\xi\,\theta_{2}\right)^{\frac{\nu-1}{\nu+1}}\,s\,,\qquad
\theta_{4}=\frac{2\,\alpha}{\theta_{2}\,\xi\,s}\,,
\label{eq:condth}
\end{equation}
where $\nu=(n-1)^{-1}$ and having also introduced the parameters $\alpha$, $\xi,$ and $s$. We can also express $\xi$ in terms of the AdS radius $L$:
\begin{equation}
\xi=\frac{2\,L\,\nu}{-1+\nu}\,\frac{1}{\sqrt{1-\alpha^{2}\,L^{2}}}\,.
\end{equation}
The truncation to the dilaton $\varphi$ is consistent provided equation \eqref{eq:construnc} is satisfied. This relation requires, in the new parametrization \eqref{eq:condth}, the condition
\begin{equation}
(s^2-1)\,(\nu^2-1)\,\alpha\,\sqrt{1-L^2\,\alpha^2}\=0\;,
\end{equation}
which is solved, excluding values $n=0$ and $n=2$, either for pure electric FI terms ($\alpha=0$) or for $s=\pm1$.\par
%
%
After the shift \eqref{eq:phishift}, the scalar field $z$ is expressed as
\begin{equation}
z\=\left(\theta_{2}\,\xi\right)^{-\frac{2\nu}{\nu+1}}\,e^{\lambda\,\varphi}\,, \end{equation}
and the same redefinition for the potential (in the general case $s=\pm1$) yields
\begin{equation}\label{eq:Vpotalphanu}
\begin{split}
V(\varphi)~=\,&-\frac{\alpha^2}{\nu^2}\,\left(\frac{(\nu-1)(\nu-2)}{2}\,
    e^{-\varphi\,\ell\,(\nu+1)} + 2\,(\nu^2-1)\,e^{-\varphi\,\ell} +\frac{(\nu+1)(\nu+2)}{2}\,e^{\varphi\,\ell\,(\nu-1)}\right)+
\\
    &+\frac{\alpha^2-L^{-2}}{\nu^2}\,\left(\frac{(\nu-1)(\nu-2)}{2}\,
    e^{\varphi\,\ell\,(\nu+1)} + 2\,(\nu^2-1)\,e^{\varphi\,\ell} +\frac{(\nu+1)(\nu+2)}{2}\,e^{-\varphi\,\ell\,(\nu-1)}\right)
\end{split}
\end{equation}
where
\begin{equation}
\ell=\frac{\lambda}{\nu}\:,\qquad\quad \lambda=\sqrt{\frac{2\,\nu^2}{-1+\nu^2}}\:,
\label{eq:ell}
\end{equation}
and having disposed of $\theta_2$ by redefinitions \eqref{eq:condth}.\par\smallskip
Let us now rewrite $\varphi$ as
\begin{equation}
\varphi\=\frac{\log(x)}{\ell}\=\sqrt{\frac{-1+\nu^2}{2}}\,\log(x)\:,
\label{eq:phi-x}
\end{equation}
so that
\begin{equation}
z\=x^\nu\left(\theta_2\,\xi\right)^{-\frac{2\,\nu}{1+\nu}}\:,
\label{eq:z-x}
\end{equation}
and the scalar potential is now expressed as:
\begin{equation}\label{eq:Vx}
\begin{split}
V(x)\=
    \frac{1}{2\,\nu^2\,L^2}\Big(&x\left(1-\alpha^2\,L^2\right)\big(4-4\,\nu^2-x^\nu(-2+\nu)(-1+\nu)-x^{-\nu}(1+\nu)(2+\nu)\big)-
\\
        &-\alpha^2\,L^2\;x^{-1-\nu}\big(2-3\,\nu+\nu^2+x^{2\,\nu}(1+\nu)(2+\nu)+4\,x^\nu(-1+\nu^2)\big)\Big)\:,
\end{split}
\end{equation}
The complex superpotential $\mathcal{W}$ can be obtained from \eqref{eq:W} and in this new parametrization reads
\begin{equation}
\mathcal{W}(x)\=
    \frac{\sqrt{1-\alpha^2L^2}}{2\,L\,\nu}\;x^{\frac{1-\nu}{2}}\,\big(1+\nu+x^\nu(-1+\nu)\big)
    \:+\:i\;\frac{\alpha}{2\,\nu}\,x^{\frac{-1-\nu}{2}}\,\big(1-\nu-x^\nu(1+\nu)\big)\:.
\label{eq:Wx}
\end{equation}

\subsubsection[\texorpdfstring{Case $\alpha=0$}{}]%
{\boldmath Case $\alpha=0$ \unboldmath}
\label{subsubsec:alpha0}
Some things change in the $\alpha=0$ configuration. Once defined $\mathcalboondox{W}_0=\mathcalboondox{W}\big|_{\alpha=0}$, in this case one finds that
\begin{equation}
\partial_\chi\mathcalboondox{W}_0\big\rvert_{{}_{\chi=0}}=\:0\;,
\end{equation}
unlike the previous general \eqref{eq:nonconsistentWre}, so that now the truncation becomes consistent also at the level of the superpotential. In particular, one finds that the imaginary part of the truncated complex superpotential vanishes, being proportional to $\alpha$ (see also \eqref{eq:Wx}).\par
The dilaton-truncated scalar potential has now the form
\begin{equation}
\begin{split}
V(\varphi)\:=\,-\,\frac{1}{L^2\,\nu^2}\,\left(\frac{(\nu-1)(\nu-2)}{2}\,
    e^{\varphi\,\ell\,(\nu+1)} + 2\,(\nu^2-1)\,e^{\varphi\,\ell} +\frac{(\nu+1)(\nu+2)}{2}\,e^{-\varphi\,\ell\,(\nu-1)}\right),
\end{split}
\end{equation}
that, in terms of reparametrization \eqref{eq:phi-x}, reads
\begin{equation}
V_0(x)\=\frac{x}{2\,L^2\,\nu^2}\,\big(4-(\nu+1)(\nu+2)\,x^{-\nu}-(\nu-2)(\nu-1)\,x^{\nu}-4\,\nu^2\big)\;,
\end{equation}
and can be expressed through \eqref{eq:VWre} using the following expression for the real superpotential
\begin{equation}
\mathcalboondox{W}_0(x)\=\frac{x^{\frac{1-\nu}{2}}}{2\,L\,\nu}\,\big(1+\nu+x^\nu(-1+\nu)\big)\;,
\end{equation}
that coincides with the (former complex) superpotential \eqref{eq:Wx} in the $\alpha=0$ case.

\subsection[\texorpdfstring{$\mathcal{N}=2$ model and \,$\mathcal{N}=8$ truncations}{}]%
{\boldmath $\mathcal{N}=2$ model and \,$\mathcal{N}=8$ truncations \unboldmath}
\label{sec:trunc}
The original ${\rm SO}(8)$ gauging of the maximal $\mathcal{N}=8$, $D=4$ supergravity \cite{deWit:1981sst,deWit:1982bul} and its generalizations to non-compact/non-semisimple ${\rm CSO}(p,q,r)$ gauge groups, $p+q+r=8$ \cite{Hull:1984yy,Hull:1984vg} are part of a broader class of gauged maximal theories, usually referred to as ``dyonic'' gaugings \cite{DallAgata:2011aa,DallAgata:2012mfj,DallAgata:2012plb,DallAgata:2014tph,Inverso:2015viq}. The construction of the latter is performed by exploiting the freedom in the initial choice of the symplectic frame in the maximal theory: different frames can be in fact obtained by rotating the original one \cite{deWit:1981sst} by a suitable symplectic matrix. If we decide to gauge the same ${\rm SO}(p,q)$ group, $p+q=8$, in different symplectic frames, a one-parameter class of inequivalent theories featuring the same gauge group ${\rm SO}(p,q)$ can be constructed. The latter are named \emph{$\omega$-deformed} ${\rm SO}(p,q)$ models, $\omega$ being the angular variable parameterizing the chosen frame.\footnote{%
these deformed theories feature a richer vacuum structure than the original model \cite{deWit:1981sst,deWit:1982bul,Hull:1984yy,Hull:1984vg}, corresponding to the $\omega=0$ value.
}\par
The truncated supergravity action explicitly reads
\begin{equation}
I=-\frac{1}{8\pi G} \,\int_{M}d^{4}x\;\sqrt{-g}\,\left(\frac{R}{2}-\frac{1}{2} \left(\partial\varphi\right)^2+V(\varphi)\right) \;.
\end{equation}
The (infinitely many) theories we have introduced in this Section contain all the possible one-dilaton consistent truncations of the $\omega$-deformed ${\rm SO}(8)$ gauged maximal supergravities. Indeed, if we perform the change of variables
\begin{equation}
\alpha\=L^{-1}\sin(\omega)\:,
\end{equation}
the $\mathcal{N}=2$ scalar field potential with electromagnetic gauging results in
\begin{equation}
V(\varphi)\=\cos^{2}(\omega)\:\mathcal{Q}(\varphi)\:+\:\sin^{2}(\omega)\:\mathcal{Q}(-\varphi)\:,
\end{equation}
where
\begin{equation}
\mathcal{Q}(\varphi)=-\frac{L^{-2}}{\nu^{2}}\,\left(\frac{(\nu -1)(\nu-2)}{2}%
\,e^{\varphi \,\ell \,(\nu +1)}+2\,(\nu ^{2}-1)\,e^{\varphi \,\ell }+\frac{1}{2}
\,(\nu +1)(\nu +2)\,e^{-\varphi \,\ell \,(\nu -1)}\right)\,,
\end{equation}
$\ell$ being defined in \eqref{eq:ell}. The gauging is then purely electric if $\omega=0$ and purely magnetic when $\omega =\pi/2$, while self-duality invariance of the potential can be then expressed as
\begin{equation}
\omega \;\rightarrow \;\omega +\frac{\pi }{2}\:,\qquad
\varphi\;\rightarrow \;-\varphi \:.
\end{equation}
The truncations can be characterized by the breaking of the $\mathrm{SO}(8)$
gauge group to the following stabilizers of the dilatonic field $\varphi$
\begin{equation}
\begin{tabular}{rcl}
$\nu =\tfrac{4}{3}$ & $\rightarrow $ & $\mathrm{SO}(7)$ \ , \\[1ex]
$\nu =2$ & $\rightarrow $ & $\mathrm{SO}(6)\times \mathrm{SO}(2)$ \ , \\[1ex]
$\nu =4$ & $\rightarrow $ & $\mathrm{SO}(5)\times \mathrm{SO}(3)$ \ , \\[1ex]
$\nu =\infty $ & $\rightarrow $ & $\mathrm{SO}(4)\times \mathrm{SO}(4)$ \ .%
\end{tabular}%
\end{equation}
When \,$\nu=\infty$\, or \,$\nu=2$, one must also set $\omega =0$ to have an embedding in $\mathcal{N}=8$ supergravity: this allows to consistently uplift our solutions to corresponding $\omega $-rotated models. For a more detailed analysis about the embedding of our models within maximal four-dimensional supergravity, we refer to \cite{Anabalon:2020pez}.


\section{Hairy BH solutions in AdS-CFT duality}\label{sec:HairyBH}
In  this  section,  we  present  a  general  family  of  exact  asymptotically AdS neutral hairy black hole solutions, with the non-trivial dilaton potential obtained in Section \ref{sec:Sugra}.  We use the quasilocal formalism  of  Brown  and  York  \cite{Brown:1992br}, supplemented  with  counterterms to  study  their thermodynamics. We compute the quasilocal stress tensor, energy, on-shell  Euclidean action (together with the corresponding thermodynamic potential) and show that the first law of thermodynamics and quantum statistical relation are satisfied. These hairy solutions have a dual interpretation as triple-trace deformations in field theory.

\subsection{Hairy BH  as a triple-trace deformation}
We are interested in a particular case of \cite{Anabalon:2017yhv}, namely
planar hairy black holes in the limit $\alpha=0$. The general metric ansatz
is
\begin{equation}
\label{metricansatz}
ds^{2}=\Upsilon(x)\left(f(x)\,dt^{2}-\frac{\eta^{2}dx^{2}}{f(x)}-d\Sigma\right)\,, \qquad\quad d\Sigma=dy^2+dz^2\:,\qquad
\end{equation}
and we choose the following conformal factor:
\begin{equation}
\Upsilon(x)~=\frac{L^2\,\nu^2\,x^{\nu-1}}{\eta^2\,(x^\nu-1)^2}\;,
\end{equation}
so that the equation of motion for the dilaton can be easily integrated.\par
It was shown in \cite{Anabalon:2017yhv} that there exist two distinct
families of solutions, each one of them containing two branches. When the
horizon topology is toroidal, the first family is characterized by the
following dilaton and metric function:
\begin{equation}
\begin{split}  \label{f1}
\varphi(x)=-\ell^{-1}\,\ln(x)\;, \qquad f(x)= 1+\alpha^{2}L^2\,\left(-1+%
\frac{x^2}{\nu^2}\,\Big((\nu+2)\,x^{-\nu}-(\nu-2)\,x^\nu+\nu^2-4\Big)%
\right)\,.
\end{split}
\end{equation}
The second family of planar hairy black holes can be obtained from a
symmetry transformation of the action, namely $\varphi \rightarrow -\varphi$
and $\alpha^2 \rightarrow L^{-2}-\alpha^2$, and so the new expressions for
the dilaton and metric function are:
\begin{equation}  \label{f2}
\begin{split}
\varphi(x)=\ell^{-1}\,\ln(x)\;, \qquad f(x)= 1+(1-\alpha^{2}L^2)\,\left(-1+%
\frac{x^2}{\nu^2}\,\Big((\nu+2)\,x^{-\nu}-(\nu-2)\,x^\nu+\nu^2-4\Big)%
\right)\,.
\end{split}%
\end{equation}
The $x$ coordinate is not the canonical radial coordinate of AdS spacetime.
One can easily check that the metric is asymptotically AdS and the conformal
boundary is located at $x=1$, which corresponds to $\varphi=0$. One can compute the Ricci scalar to find the location of the singularity; however, the same
information can be straightforwardly obtained from the dilaton's profile. We
observe that the dilaton diverges when $x=0$ and $x=+\infty$ and so there
exist two disconnected branches for each family, one in the range $x\in[0,1)$
and the other in the range $x\in(1,\infty]$.\par\smallskip
In what follows, we consider the case $\alpha=0$ and only the branches with
$\varphi >0$. In this case, the superpotential is real and, as we are going
to prove now, there exist regular hairy black holes pertaining to the second
family. Explicitly, when $\alpha=0$, the metric function of the first family \eqref{f1} becomes trivial, $f(x)=1$, but the metric function of the second family \eqref{f2} gives a non-trivial function of $x$,
\begin{equation}  \label{regular}
f(x)\=\frac{x^2}{\nu^2}\,\Big((\nu+2)\,x^{-\nu}-(\nu-2)\,x^\nu+\nu^2-4\Big)\,.
\end{equation}
This function vanishes when
\begin{equation}
\label{eqhor}
f(x_\text{h})=0 \quad\Longrightarrow\quad (x_\text{h})^{\nu}=\frac{\nu^{2}-4\pm\nu\sqrt{\nu^{2}-4}}{\nu-2}\:,
\end{equation}
where $x_\text{h}$ indicates the horizon location, that exists only for $x>1$ and $\nu>2$ (the positive branch of the second family). For this specific case, we are
going to construct the thermal superpotential and an exact $\beta$-function.

To obtain the boundary conditions for the dilaton, let us now use the canonical coordinates in AdS with the ansatz
\begin{equation}
ds^{2}\=g_{tt}(r)\,dt^{2}+g_{rr}(r)\,dr^{2}
    -\left(\frac{r^2}{L^2} + O\left(r^{-2}\right)\right)d\Sigma\:.
\end{equation}
We then have
\begin{equation}
\Upsilon(x)\=\frac{r^2}{L^2} + O\left(r^{-2}\right)\:,
\end{equation}
from which we can get asymptotically%
\footnote{%
for $\nu=3$, one can obtain an exact expression, but in general the change
of coordinates can be obtained only perturbatively.}
the following expansion of the coordinate $x$ in terms of the canonical radial coordinate $r$ of AdS:
\begin{equation}  \label{CC}
x \= 1 + \left(\frac{L^2}{r\,\eta}+L^6\,\frac{1-\nu^2}{24\,\left(r\,\eta%
\right)^3}\right) +L^8\,\frac{\nu^2-1}{24\,\left(r\,\eta\right)^4}%
+O\left(r^{-5}\right)\:.
\end{equation}
With this change of coordinates, we get the usual fall-off of the scalar
field in AdS:
\begin{equation}  \label{fall off}
\varphi\=L^2\,\frac{\varphi_0}{r}+L^4\,\frac{\varphi_1}{r^2}%
+O\left(r^{-3}\right)\= L^2\,\frac{1}{\ell\,\eta\,r}-L^4\,\frac{1}{%
2\,\ell\,\eta^2\,r^2}+ O\left(r^{-3}\right)\:.
\end{equation}
The Breitenlohner-Freedman (BF) bound \cite{Breitenlohner:1982bm} in four dimensions is $m_{\textsc{bf}}^{2}=-\tfrac{9}{4\,L^{2}}$. The conformal mass of the dilaton is given by \,$m^{2}=-2/L^{2}$, with
\begin{equation}
\Delta_{\pm}=\frac{3}{2}\pm\sqrt{\frac{9}{4}+m^{2}L^{2}}
\quad\;\Longrightarrow \quad\;
\Delta_{+}=2\,,\quad \Delta_{-}=1\:,
\end{equation}
and so
\begin{equation}
\frac{1}{4\,L^{2}}\,>\,m^{2}\,\geq\,-\frac{9}{4\,L^{2}}\:.
\label{bf}
\end{equation}
Therefore, both modes in the dilaton's fall-off are normalizable, and they are related by the following boundary conditions:
\begin{equation}
\label{boundarydeformation}
\varphi_0=\frac{1}{\ell\,\eta}\:, \qquad
\varphi_1=-\frac{\ell}{2}\,\varphi_{0}^2\:, \qquad w(\varphi)=-\frac{\ell}{6}\,\varphi_{0}^{3}\:.
\end{equation}
It can be then explicitly checked that these boundary conditions preserve
the isometries of AdS \cite{Hertog:2004dr,Henneaux:2004zi,Anabalon:2015xvl}:
\begin{equation}
\varphi_{1}=\frac{d w}{d\varphi_{0}}\:, \qquad
w(\varphi_{0})-\frac{\varphi_{0}\varphi_{1}}{3}=0\:.
\end{equation}
In the context of AdS-CFT duality, Witten has interpreted the general mixed boundary conditions for the scalar field as a multi–trace deformation of the dual CFT of the form $\int\!d^{3}x\:w\left[\mathcal{O}(x)\right]$\, \cite{Witten:2001ua}. In our case, the boundary perturbation \eqref{boundarydeformation} generates a triple-trace deformation of the dual field theory \cite{Hertog:2004dr,Witten:2001ua}:%
\footnote{%
see, also, \cite{Freedman:2016yue} for a detailed discussion of the cubic coupling in ABJM theory \cite{Aharony:2008ug}
}
\begin{equation}
I_\textsc{cft} \;\longrightarrow \; I_\textsc{cft}+\frac{\ell}{6}\int{d^{3}x\:\mathcal{O}^3}\;.
\end{equation}
It is important to notice that the parameter $\nu$ affects the coupling of
deformation through the parameter $\ell$, but, regardless of the value of $\nu$, the dual field theory is always deformed by a triple-trace deformation.\par
Since the boundary conditions are such that the conformal symmetry in the
boundary is preserved, the ADM mass \cite{Ashtekar:1984zz,Ashtekar:1999jx}
matches the holographic mass \cite{Anabalon:2015xvl,Anabalon:2014fla} and
can be read off from the expansion of the metric:
\begin{equation}
g_{tt}\=\Upsilon(x)\,f(x)\=
    \frac{r^2}{L^2}-\frac{\nu^2-4}{3\,\eta^3}\,\frac{L^4}{r}+O\left(r^{-2}\right)\:.
\end{equation}
For convenience, we consider $\eta>0$ and so the mass density of the planar
hairy black hole is
\begin{equation}
\rho = \frac{M}{\sigma}=L^4\,\frac{\mu}{8\pi G}\;, \qquad\quad
\mu \equiv \frac{\nu^2-4}{3\,\eta^3}\:,
\end{equation}
with $\sigma=L^{-2}\int d\Sigma$, defined to be dimensionless. We notice that the mass density is positive defined when $\nu > 2$, a result compatible with the one obtained from the condition of existence of the horizon (\ref{eqhor}).


\subsection{Holographic stress tensor}\label{holstress}
To confirm that, indeed, the ADM and holographic mass match, let us compute the holographic stress tensor using the counterterms proposed in \cite{Anabalon:2014fla}:
\begin{equation}
\tau_{ab}\=\frac{1}{8\pi G}\left(K_{ab}-h_{ab}\,K+\frac{2}{L}h_{ab}
    -L\,G_{ab}\right)+\frac{h_{ab}}{8\pi G}
    \left(\frac{\varphi^{2}}{2L}-\frac{\ell\,\varphi^{3}}{6L}\right)\,.
\end{equation}
We choose the foliation $x=\text{const.}$, with the induced metric $h_{ab}$ on each $3$ -dimensional hypersurface, for which the boundary is at $x=1$.
Here, $G_{ab}$ is the Einstein tensor for the foliation and $K_{ab}$ is the extrinsic curvature. The trace of the induced metric is denoted by $h$ and the one of the extrinsic curvature by $K$.\par
For the hairy BH solution (\ref{regular}), we get
\begin{equation}
\tau_{tt}=\frac{L}{R}~\frac{\mu\,L^{4}}{8\pi G}\,, \qquad\;
\tau_{yy}=\tau_{zz}=\frac{L}{R}~\frac{\mu\,L^{4}}{16\pi G}\:.
\end{equation}
The geometry where the dual field theory lives is related to the induced metric on the boundary by a conformal factor,
\begin{equation}
ds^{2}_\text{bound}=\frac{R^{2}}{L^{2}}\left(dt^{2}-d\Sigma\right), \qquad\quad
ds^{2}_\text{dual}=\gamma_{ab}\,dx^{a}dx^{b}=dt^{2}-d\Sigma\:.
\end{equation}
Consequently, the quasilocal stress tensor on the gravity side and the one of the dual field theory are related by \cite{Myers:1999psa}
\begin{equation}
\langle\tau_{ab}^{\text{dual}}\rangle=\lim_{R\rightarrow\infty}\frac{R}{L}\,\tau_{ab}\:,
\end{equation}
and so, we obtain
\begin{equation}
\langle\tau^{\text{dual}}_{ab}\rangle\=\frac{L^{4}\mu}{16\pi G}\left(3\,\delta_{a}^{0}\,\delta_{b}^{0}-\gamma_{ab}\right)\=
(\rho+p)\,u_{a}\,u_{b}-p\,\gamma_{ab}\:.
\end{equation}
For observers on the boundary (with $u_{a}=\delta_{a}^{0}$), we explicitly obtain
\begin{equation}
\label{fluid}
\rho+p=\frac{3L^{4}\mu}{16\pi G}\,, \qquad p=\frac{L^{4}\mu}{16\pi G} \;\;\quad\Longrightarrow\quad\;\;
\rho=\frac{L^{4}\mu}{8\pi G}\:,
\end{equation}
that corresponds to the one of a conformal gas with energy density $\rho$ and pressure $p$. The holographic stress tensor is covariantly conserved and its trace vanishes, $\langle\tau^{\text{dual}}\rangle=0$, as expected for boundary conditions of the dilaton that preserve the conformal symmetry.

The conserved charges can be obtained from the quasilocal formalism of Brown and York \cite{Brown:1992br}. For the Killing vector $\xi^{j}=\partial/ \partial t$, the conserved quantity is the total energy of the black hole (including the hair)
\begin{equation}
E\:=\int{d\Sigma^{i}\tau_{ij}\xi^{j}}\=L^{4}\frac{\sigma\mu}{8\pi G}\:,
\end{equation}
where $d\Sigma^{i}$ is the planar surface at infinity with $t=\text{const}$.
The same result was obtained in (\ref{fluid}) by using the stress tensor of the dual conformal gas with energy density $\rho$ and pressure $p$.


\subsection{Quantum statistical relation}
In this section we closely follow \cite{Anabalon:2015xvl}%
\footnote{The advantage of this method compared with the usual holographic renormalization for the general solutions (\ref{f1}) and (\ref{f2}) is that it can be used even for a complex superpotential \cite{Anabalon:2017yhv}, though in this work when $\alpha=0$ the superpotential is real. However, we are going to compare the two methods in Section \ref{holoren}.}
and consider the  regularized Euclidean action
\begin{equation}
\label{action}
I^{\textsc{e}}\left[g^{\textsc{e}},\varphi\right]\=
\frac{1}{8\pi G}\int_{M}d^{4}x\:\sqrt{g^{\textsc{e}}}\left( \frac{R}{2}-\frac{1}{2}\,\left(\partial\varphi\right)^{2}+V(\varphi)\right)
\:-\: \frac{1}{8\pi G}\,\int_{\partial M}\!\!d^{3}x\:\sqrt{h}\:K \:+\: I_\mathrm{ct}^{\textsc{e}} \:+\: I_{\varphi}^{\textsc{e}}\:,
\end{equation}
with the gravitational counterterm \cite{Balasubramanian:1999re}
\begin{equation}
\begin{split}
I_\mathrm{ct}^{\textsc{e}}\=\frac{1}{8\pi G}\,\int_{\partial M} \!\!d^{3}x\:\sqrt{h}\,\left(
\frac{2}{L}+\frac{L}{2}\,R(h)\right)\,.
\end{split}
\end{equation}
Since we would like to compare this method with the one using the superpotential as a counterterm, we write in detail the contribution in the boundary of the first three terms in the action:
\begin{equation}
I^{\textsc{e}}_\text{bulk}+I^{\textsc{e}}_{\textsc{gh}}+I_\text{ct}^{\textsc{e}}\=
    -\frac{\mathcal{A}}{4G}+\frac{\sigma L^{2}}{8\pi G\,T}\left(L^{2}\mu+\frac{2\,\ell\,L^{2}}{3}\,\varphi_{0}^{3}
    -\frac{r\,\varphi_{0}^{2}}{2}\right)\,,
\end{equation}
%
where the temperature and entropy are
\begin{equation}
T=\frac{1}{4\pi\eta}\frac{(\nu^{2}-4)\,(x_\text{h}^{\nu}-1)^{2}}{\nu^{2}\,x_\text{h}^{\nu-1}}\,, \qquad S=\frac{\sigma\,L^2\,\Upsilon(x_\text{h})}{4G}=\frac{\mathcal{A}}{4G}\:.
\end{equation}
We can identify a divergence in the last term and, to regularize the action, it is necessary to include the scalar field counterterms,
\begin{equation}
I_{\varphi}^{\textsc{e}}\=\frac{1}{8\pi G}\int_{\partial M}\!\!d^{3}x\:\sqrt{h}
\,\left(\frac{\varphi^{2}}{2\,L}+\frac{w(\varphi_{0})}{L\,\varphi_{0}^{3}}\,\varphi^{3}\right)
\=\frac{1}{8\pi G}\int_{\partial M}\!\!d^{3}x\:\sqrt{h}
\,\left(\frac{\varphi^{2}}{2\,L}-\frac{\ell\,\varphi^{3}}{6\,L}\right)\:,
\end{equation}
that, after integration at the boundary, give
\begin{equation}
I_{\varphi}^{\textsc{e}}\=
\frac{\sigma\,L^{2}}{8\pi G\,T}\left(-\frac{2\,\ell\,L^{2}}{3}\,\varphi_{0}^{3} +\frac{r\,\varphi_{0}^{2}}{2}\right)\:.
\end{equation}
The on-shell finite action $I^{\textsc{e}}=I^{\textsc{e}}_\text{bulk}+I^{\textsc{e}}_{\textsc{gh}}+I_\text{ct}^{\textsc{e}}+I^{\textsc{e}}_{\varphi}$\, is related to the free energy and the quantum statistical relation is satisfied:
\begin{equation}
F\=I^{\textsc{e}}\,T=M-T\,S\,, \qquad\quad M=L^{4}\,\frac{\sigma\,\mu}{8\pi G}\:.
\end{equation}
One can also explicitly verify that the first law is satisfied by doing the variation with respect to the integration constant $\eta$ (the other parameter, $\nu$, is a parameter of the theory). As a final observation, we notice that $\sigma$ is, in principle, a surface with infinite area. One should then work with `densities' or make identifications in the geometry to obtain a toroidal surface with finite area.


\section{Holographic applications}\label{sec:Holography}
In this section we are going to construct the thermal superpotential for the exact regular hairy black hole solution (\ref{regular}). We start with a brief review of \cite{Gursoy:2008za} and explicitly show how the second order equations can be rewritten as first order equations, when the thermal superpotential is introduced. Then, we use the thermal superpotential to investigate the holography of the hairy black hole and the corresponding RG flow. We are also going to present a domain wall/black hole duality in $\omega$-deformed supergravity.

\subsection{Thermal superpotential and holographic renormalization}
\label{holoren}
Before presenting a detailed analysis of hairy BHs in extended supergravity with the ansatz (\ref{metricansatz}), let us discuss the `domain wall' coordinate ansatz:
\begin{equation}
\label{ansatzRG}	ds^{2}\=e^{2A(u)}\left(e^{g(u)}\,dt^{2}-dy^{2}-dz^{2}\right)-\frac{du^{2}}{e^{g(u)}}\:,
\end{equation}
that corresponds to a domain wall when $g(u)=0$ and to a planar black hole for $g(u)\neq0$. In these coordinates, the holographic interpretations can be stated unambiguously.\par
The Einstein equations
\begin{equation}
E_{\mu \nu }\:\equiv\:
G_{\mu \nu }-\partial _{\mu }\varphi\,\partial_{\nu }\varphi
+g_{{\mu \nu }}\left(\frac{1}{2}\left(\partial \varphi\right)^{2}
    -V(\varphi)\right)\=0\;,
\end{equation}
using ansatz (\ref{ansatzRG}) become (here, the derivative with respect to $u$, $d/du$, is denoted by $^{\prime}$)
\begin{equation}
\begin{split}
E_{t}^{t}-E_{u}^{u}&=0  \quad\Leftrightarrow\quad
2\,A^{\prime\prime}+{\varphi^\prime}^2=0\:,
\\[\jot]
E_{u}^{u}&=0  \quad\Leftrightarrow\quad
3\,{A^\prime}^2-\frac{1}{2}\,{\varphi^\prime}^2+A^\prime\,g^\prime+e^{-g}\,V=0\:,
\\[\jot]
E_{y}^{y}-E_{u}^{u}&=0  \quad\Leftrightarrow\quad
{g^\prime}^2+g^{\prime\prime}+3\,A^\prime\,g^\prime=0\:.
\end{split}
\end{equation}
The first equation involves just the dilaton and the warp factor $A(u)$ and, since the metric function $g(u)$ does not appear explicitly in its
expression, this equation is the same for the domain wall and planar black
hole. This important feature hints to the fact that a generalization of the
superpotential at finite temperature should use directly this specific
equation, rather than the usual relation between the potential and
superpotential in supergravity. We transform this second order equation in
two first order equations by defining the thermal superpotential $W(\varphi)$ as
\begin{equation}
\label{defsuper}
2\,A^{\prime\prime}+{\varphi^\prime}^2\=0
\;\;\quad\Leftrightarrow\quad\;\;
\varphi^\prime(u)=\frac{dW(\varphi)}{d\varphi} \,,
\quad
A^\prime(u)=-\frac{1}{2}\,W(\varphi)\;.
\end{equation}
Then, the second and third Einstein equations can be rewritten in terms of
the superpotential as
\begin{equation}
\frac{1}{2}\left( \frac{dW}{d\varphi }\right)^{2}
-\frac{3}{4}\,W^{2}+\frac{1}{2}\,g^\prime\,W\=e^{-g}\,V\:,
\qquad
{g^\prime}^2+g^{\prime\prime}-\frac{3}{2}\,g^\prime\,W\=0\;.
\end{equation}
When $g(u)=0$, we recover the usual relation between the potential and
superpotential from the (fake) supergravity. However, we observe that at
finite temperature the non-trivial function $g(u)$ plays an important role
in obtaining the thermal superpotential.

To apply this formalism to the hairy black hole solution (\ref{regular}), we have to use the $x$-coordinate. In these new coordinates, the first Einstein equation and thermal superpotential become
\begin{equation}
\frac{\varphi^{\prime}}{\eta\,\sqrt{\Upsilon}}=\frac{dW}{d\varphi}\,, \qquad\quad
W(\varphi)=-\frac{\Upsilon^{\prime}}{\eta\,\Upsilon^{3/2}}\:.
\label{ec2}
\end{equation}
Using now the other two Einstein equations, we obtain the relation between the potential and superpotential in the $x$-coordinate system as
\begin{equation}
V=\left(\left(\frac{dW}{d\varphi}\right)^{2}-\frac{3}{2}\,W^{2}\right)\frac{f}{2}
    +\frac{W\,f^{\prime}}{2\,\eta\,\sqrt{\Upsilon}}\:.
\end{equation}
For the exact solution (\ref{regular}), this equation can be integrated and the resulting thermal superpotential is
\begin{equation}
W(\varphi)\=
\frac{1}{\nu\,L}\left((\nu+1)\,e^{\varphi\,\ell\,\left(\frac{\nu-1}{2}\right)}+
(\nu-1)\,e^{-\varphi\,\ell\,\left(\frac{\nu+1}{2}\right)}\right)\:.
\label{superpotential}
\end{equation}
The relation between the supergravity real superpotential $\mathcalboondox{W}_0(\varphi)$ defined in subsection \ref{subsubsec:alpha0} and the above thermal superpotential is given by $W(\varphi)=2\,\mathcalboondox{W}_0(-\varphi)$. At first sight, this specific relation may come as a surprise. However, this can be traced back to the duality between the two families of solutions (see the expressions \eqref{f1} and \eqref{f2} for the scalar field) that changes the sign of the scalar field. The metric function is drastically changed, but only the scalar field is relevant for the superpotential.\par
As a first application in AdS-CFT duality, let us use the thermal superpotential to compute the Euclidean action. In Section \ref{holstress} we have used gravitational and dilaton counterterms to regularize the action, this method being compatible with a well defined variational principle for general mixed boundary conditions for the dilaton \cite{Anabalon:2015xvl}. Alternatively, one can use the superpotential as a counterterm to regularize the action \cite{Batrachenko:2004fd,Papadimitriou:2007sj}. The main difference is that, in this case, we do not have to add the gravitational counterterm, the whole information being contained in the thermal superpotential. Let us now check that the two methods are equivalent. The Euclidean action contains only three terms,
\begin{equation}
I^{\textsc{e}}\left[g^{\textsc{e}},\varphi\right]\=
    I^{\textsc{e}}_\text{bulk}+I^{\textsc{e}}_{\textsc{gh}}+I_{W(\varphi)}\:,
\end{equation}
the boundary contribution from the first two terms being
\begin{equation}
I^{\textsc{e}}_\text{bulk}+I^{\textsc{e}}_{\textsc{gh}}\=
-\frac{\mathcal{A}}{4G}+\frac{\sigma\,L^{2}}{8\pi G\,T} \left(-\frac{2\,r^{3}}{L^{4}}-\frac{r\,\varphi_{0}^{2}}{2}
+\frac{2\,L^{2}\,\ell}{3}\,\varphi_{0}^{3}+2\,\mu\,L^{2}\right)\,.
\end{equation}
We identify two divergent terms proportional to $r^3$ and $r$ that can be canceled out if we use the thermal superpotential as a counterterm:
\begin{equation}
I_{W(\varphi)}\=
\frac{1}{8\pi}\int_{\partial\mathcal{M}}\!\!\!d^{3}x\;\sqrt{h^{\textsc{e}}}\;W(\varphi)
\=\frac{\sigma L^{2}}{8\pi G\,T} \left(\frac{2\,r^{3}}{L^{4}}+\frac{r\,\varphi_{0}^{2}}{2}-\frac{2\,L^{2}\,\ell}{3}\,\varphi_{0}^{3}
-\mu\,L^{2}\right)\,.
\end{equation}

\subsection{Holographic renormalization group}
\label{sec:RGflow}
\begin{figure}[!h]
\centering
\begin{subfigure}[t]{.45\linewidth}
\centering\includegraphics[width=1.\linewidth]{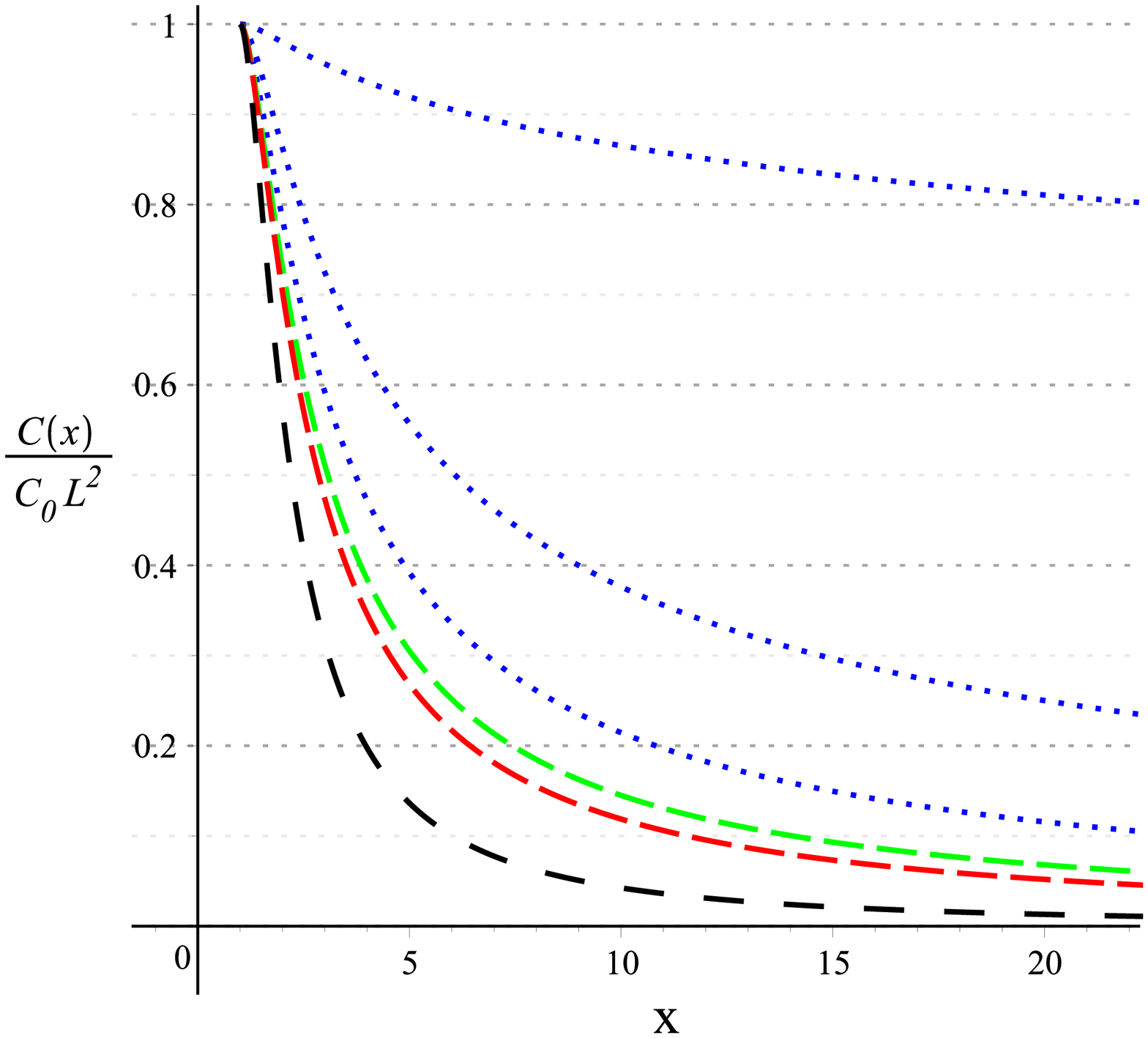}
\caption{Plot of the $c$-function as a function of the coordinate $x$, for the branch $x>1$ and $\nu$ hairy parameter values ${\nu=1.1,\,1.6,\,1.9,\,2.1,\,2.2,\,2.7}$.}
\label{subfig:monox}
\end{subfigure}
\hfill
\begin{subfigure}[t]{.52\linewidth}
\centering
\includegraphics[width=1.\linewidth]{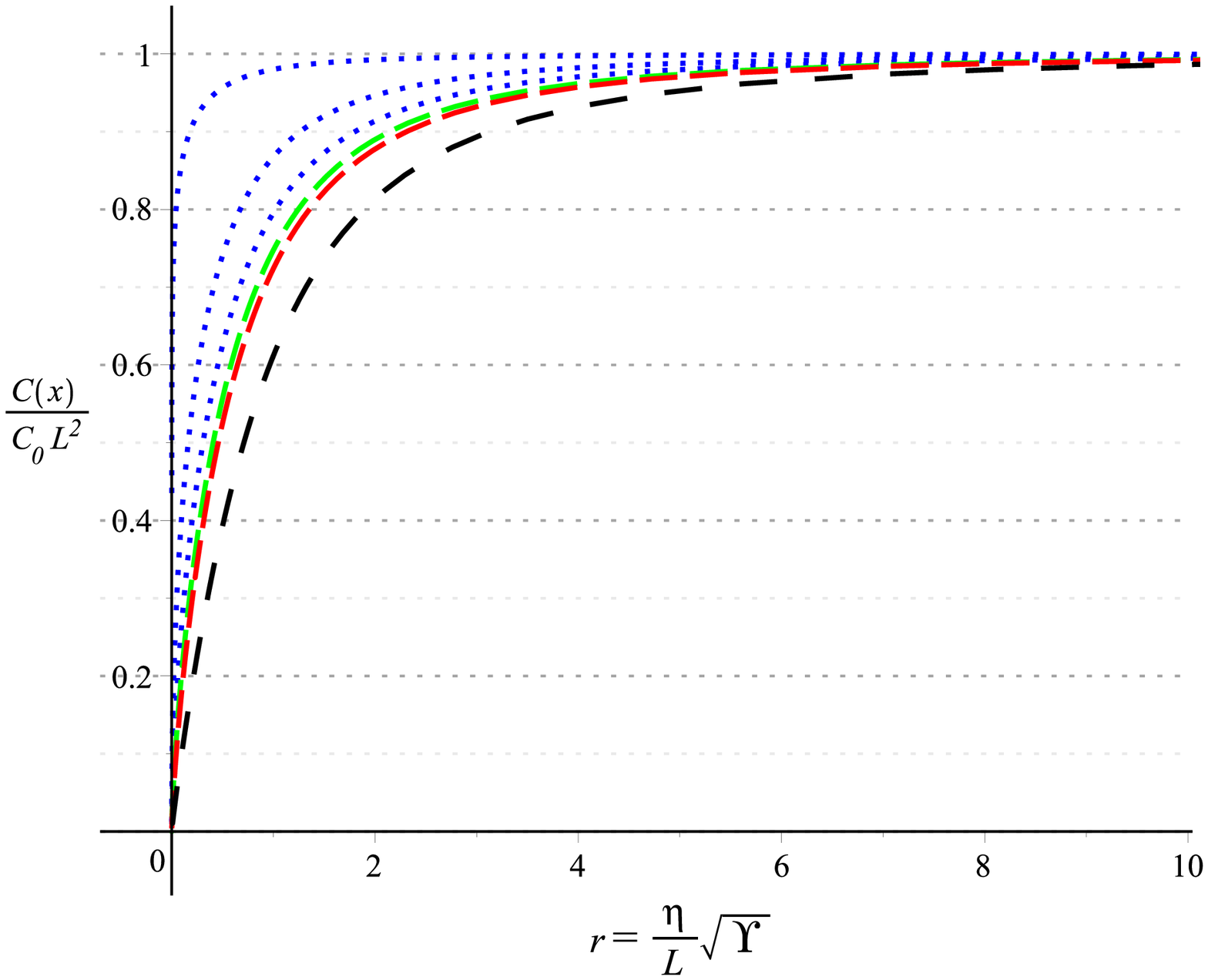}
\caption{Plot of the $c$-function in terms of the canonical radial coordinate  $r=\tfrac{\eta\,\sqrt{\Upsilon}}{L}=\tfrac{\eta}{L}\frac{\nu\,x^{\frac{\nu-1}{2}}}{x^{\nu}-1}$ for the same values of the $\nu$ parameter.}
\label{subfig:monor}
\end{subfigure}
%
\end{figure}

Once we have the thermal superpotential, we can explicitly construct the $\beta$-function. In AdS-CFT duality, the radial coordinate is interpreted as the energy of the dual field theory. Therefore, in the `domain wall' coordinate system, the $\beta$-function can be computed as:
\begin{equation}
\beta(e^{\varphi})\=\frac{\varphi^{\prime}}{A^{\prime}}\:e^{\varphi}\= -\frac{2}{W}\left(\frac{dW}{d\varphi}\right)e^{\varphi}\:.
\label{be2}
\end{equation}
Written as a function of the thermal superpotential, the $\beta$-function is in turn directly computed as a function of the dilaton.  By using (\ref{superpotential}), we obtain
\begin{equation}
\beta(e^{\varphi})\=
-\frac{2}{\ell}\:\frac{e^{\nu\,\varphi\,\ell}-1}{e^{\nu\,\varphi\,\ell}(\nu+1)+\nu-1}\;e^{\varphi}\:,
\label{bet3}
\end{equation}
that, as expected, matches the result obtained from a direct computation using the dilaton expression (\ref{f2}):
\begin{equation}
\beta(e^{\varphi})\=e^{\varphi}\,\frac{\pa_{u}\varphi}{\pa_{u}A}\=
\frac{2\,\Upsilon}{\Upsilon^{\prime}}\:\varphi^{\prime}\,e^{\varphi}\:.
\end{equation}
\sloppy
The original proposal \cite{Freedman:1999gp} for an holographic $c$-function for a domain wall  can be extended at finite temperature (see, e.g., \cite{Elvang:2007ba}). The geometrical construction is based on imposing the null energy condition%
\footnote{For any light-like vector $n^{\alpha}n_{\alpha}=0$, at every point in spacetime, the matter energy-momentum tensor obeys $T_{\alpha\beta}\,n^{\alpha}n^{\beta}\geq 0$},
which captures the positivity of local energy density on the matter sector of the theory. For a gravity theory with a scalar field and its self-interaction, the null energy is satisfied, that is ${\rho+p\sim \varphi^{\prime\,2}\geq 0}$. For the metric (\ref{ansatzRG}), we obtain
\begin{equation}
-2\,A^{\prime\prime}=\varphi^{\prime\,2}>0 \quad\Longrightarrow\quad \frac{d}{du}\left(\ln{\left(\frac{1}{A^{\prime\,2}}\right)}\right)>0\:.
\label{cfunc}
\end{equation}
Due to the holographic duality, there should exist a geometric $c$-function, $\mathcal{C}(u)\geq 0$, that is monotonically increasing from the bulk towards the boundary, $\mathcal{C^{\prime}}(u)\geq 0$, so that we have
\begin{equation}
\frac{\mathcal{C}(u)^{\prime}}{\mathcal{C}(u)}\geq 0 \quad\Longrightarrow\quad
\pa_{u}\big(\ln{\mathcal{C}(u)}\big)\geq 0\:.
\label{cfunc1}
\end{equation}
By comparing (\ref{cfunc}) with (\ref{cfunc1}), we can identify the $c$-function as
\begin{equation}
\mathcal{C}(u)\=\frac{\mathcal{C}_{0}}{A^{\prime\,2}}\:.
\end{equation}
With the change of coordinates
\begin{equation}
A(u)\,\rightarrow\,\frac{1}{2}\ln{\Upsilon(x)}\,, \qquad
g(u)\,\rightarrow\,\ln{f(x)}\,, \qquad
\frac{dx}{du}=\frac{1}{\eta\sqrt{\Upsilon}}\,,
\label{trans1}
\end{equation}
the $c$-function for the planar hairy black hole solution becomes
\begin{equation}
\mathcal{C}(x)\=\mathcal{C}_{0}\left(\frac{2\,\eta\,\Upsilon^{3/2}}{\Upsilon^{\prime}}\right)^{2} \=\mathcal{C}_{0}\,L^{2}\left(\frac{2\,\nu\,x^{\frac{\nu+1}{2}}}{\nu-1+x^{\nu}(\nu+1)}\right)^2\,.
\label{cfuncionBH}
\end{equation}
We directly confirm the monotonicity of the $c$-function (\ref{cfuncionBH}) for the hairy black hole in the plots of Fig.\ \subref{subfig:monox} and \subref{subfig:monor}.
\par
To complete the analysis, let us obtain the same result using the thermal superpotential defined in (\ref{defsuper}). We rewrite the null energy condition as
\begin{equation}
\varphi^{\prime\,2}=\left(\frac{dW}{d\varphi}\right)^{2}>0
\end{equation}
and so, analogous to equation (\ref{cfunc}), the result is
\begin{equation}
-2\,A^{\prime\prime}=\varphi^{\prime\,2}>0 \quad\Longrightarrow\quad \frac{d}{du}\left(\ln{\left(\frac{4}{W^{2}}\right)}\right)>0\:.
\end{equation}
The $c$-function can be constructed as before, and the final result
\begin{equation}
\label{cfunction}
\mathcal{C}(\varphi)\=\frac{4\,\mathcal{C}_{0}}{W^{2}}\=
    \frac{4\,L^{2}\,\mathcal{C}_{0}\,\nu^{2}\,e^{\varphi\,\ell}}{%
    \left((\nu-1)\,e^{-\varphi\,\ell\,\nu/2}+(\nu+1)
	\,e^{\varphi\,\ell\,\nu/2}\right)^{2}}
\end{equation}
matches (\ref{cfuncionBH}) when written as a function of the coordinate $x$.

\subsection{Domain wall/planar black hole duality}\label{sec:duality}
The last application we would like to present is a duality between the solutions in $\omega$-deformed $\mathrm{SO}(8)$ maximal gauged supergravity presented in Section \ref{sec:trunc}. Consider the electric (i.e.\ $\alpha=0$) potential
\begin{equation}
V_{\text{el}}\=V_{\omega=0}\=\mathcal{Q}(\varphi)
\=\frac{1}{2}\left(\frac{dW_{\text{el}}}{d\varphi}\right)^{\!2} -\frac{3}{4}\,{W_{\text{el}}}^{2}\;,
\end{equation}
where
\begin{equation}
W_{\text{el}}(\varphi)
\=\frac{1}{\nu\,L}\left((\nu+1)\,e^{-\frac{\varphi\,\ell\,(\nu-1)}{2}}+ (\nu-1)\,e^{\frac{\varphi\,\ell\,(\nu+1)}{2}}\right)\:,
\end{equation}
and, correspondingly, the magnetic (i.e.\ $\alpha=L^{-1}$) potential
\begin{equation}
V_{\text{mag}}\=V_{\omega=\frac{\pi}{2}}\=\mathcal{Q}(-\varphi )
\=\frac{1}{2}\left(\frac{dW_{\text{mag}}}{d\varphi}\right)^{\!2} -\frac{3}{4}\,{W_{\text{mag}}}^{2}\:,
\end{equation}
where
\begin{equation}
W_{\text{mag}}(\varphi)
\=\frac{1}{\nu\,L}\left((\nu+1)\,e^{\frac{\varphi\,\ell\,(\nu-1)}{2}}+ (\nu-1)\,e^{\frac{-\varphi\,\ell\,(\nu+1)}{2}}\right)
\=W_{\text{el}}(-\varphi)\:,
\end{equation}
and coincides with the one in \eqref{superpotential}. Now, let us show that in the electric frame, the magnetic
superpotential generates hairy black hole solutions. Indeed we find that,
\begin{equation}
\frac{1}{2}\left(\frac{dW_{\text{mag}}}{d\varphi}\right)^{\!2} -\frac{3}{4}\,{W_{\text{mag}}}^{2}+\frac{1}{2}\,g^{\prime}\,W_{\text{mag}}
\=e^{-g}\,V_{\text{el}}\:,
\qquad\quad {g^\prime}^{2}+g^{\prime\prime}-\frac{3}{2}\,g^\prime\,W_{\text{mag}}\=0\:,\qquad
\end{equation}
using the expressions
\begin{equation}
\begin{split}
&e^g=\frac{x^{2}}{\nu^2}\,\Big((\nu+2)\,x^{-\nu}-(\nu-2)\,x^\nu+\nu^2-4\Big)\,,\qquad\quad e^{2A}=\frac{L^{2}\nu^{2}\,x^{\nu-1}}{\eta^{2}\,(x^{\nu}-1)^{2}}\=\Upsilon(x)\,,\qquad
\\[1.5ex]
&\frac{dx}{du}=e^{-A}\,\eta^{-1}\,,\qquad\quad
\varphi=\ell^{-1}\,\ln(x)\,,
\end{split}
\end{equation}
the function $e^g$ vanishing for
\begin{equation}
(x_\text{h})^{\nu}\=\frac{\nu^{2}-4\,\pm\,\nu\,\sqrt{\nu^{2}-4}}{\nu-2}\;.
\end{equation}
Electromagnetic duality is then related to a domain wall/planar black hole duality: the electric frame superpotential generates a domain wall, while the magnetic frame superpotential generates a black hole.


\section{Discussion}\label{sec:Discussion}
We have described a supergravity framework and obtained exact neutral planar hairy black holes that, within AdS-CFT duality, can generate non-trivial RG flows in the dual field theory. While a direct study of the RG properties is an involved problem in QFT, the use of AdS-CFT duality transforms it to a much more tractable one.

In particular, we were interested in an \,$\N=2$ supergravity model featuring a single vector multiplet with a complex scalar field, whose target space (K\"{a}hler) geometry was carefully analyzed in Sect.\ \ref{sec:Sugra}. The supergravity potential of a consistent dilaton truncation of the model was explicitly rewritten in  \eqref{eq:Vpotalphanu} in terms of parameters $\alpha$, $\nu$, through suitable scalar and FI terms redefinitions. At the level of the solutions, for the general case with $\alpha\neq0$, there exist two distinct families of hairy black holes that are characterized by different boundary conditions and which are related by a symmetry of the action.\par
\begin{figure}[!h]
\centering
\begin{subfigure}[t]{.475\linewidth}
\captionsetup{skip=15pt}
\centering
\includegraphics[width=0.825\linewidth]{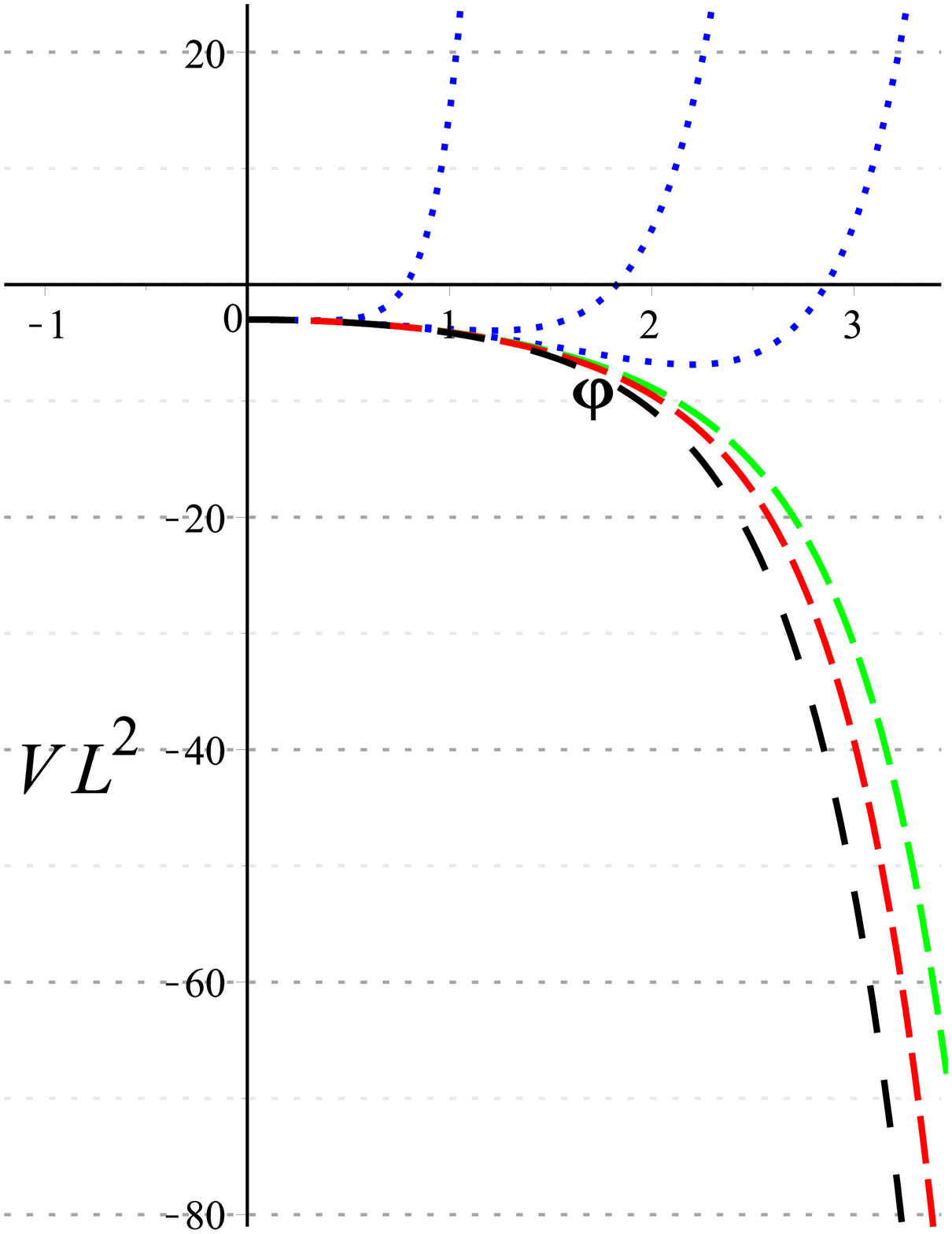}
\caption{\sloppy
Plot of the dilaton potential vs ${\varphi\geq 0}$, for  ${\nu=1.1,\,1.6,\,1.9,\,2.1,\,2.2,\,2.7}$. There are no black holes for the dot-blue curves ${\nu=1.1,\,1.6,\,1.9}$.}
\label{subfig:dilPot}
\end{subfigure}
\hfill
\begin{subfigure}[t]{.475\linewidth}
\captionsetup{skip=15pt}
\centering
\includegraphics[width=0.9\linewidth]{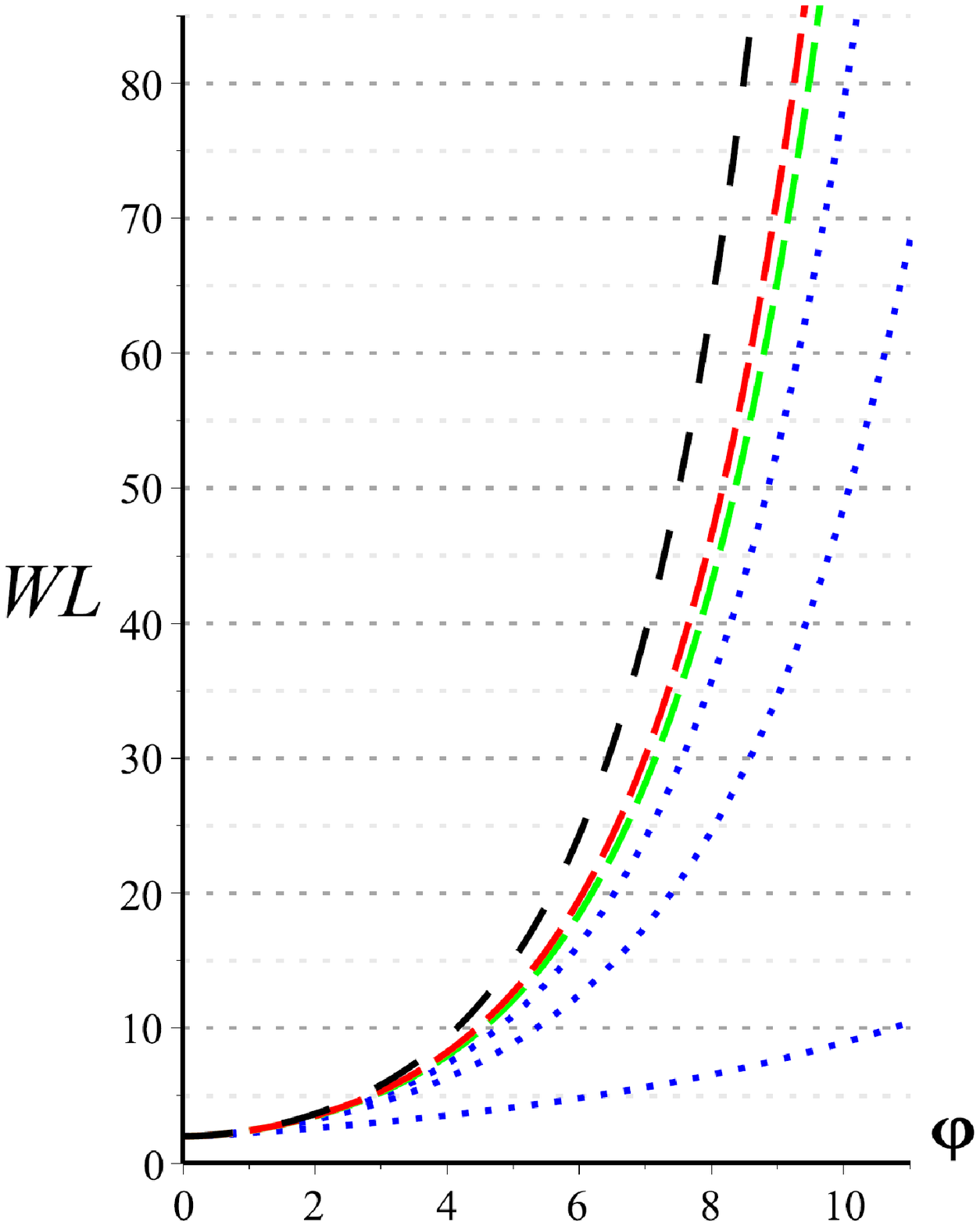}
\caption{\sloppy
Plot of the thermal superpotential vs ${\varphi\geq 0}$, for ${\nu=1.1,\,1.6,\,1.9,\,2.1,\,2.2,\,2.7}$. There are no black holes for the dot-blue curves ${\nu=1.1,\,1.6,\,1.9}$.}
\label{subfig:thermSuperpot}
\end{subfigure}
%
\end{figure}

In this work, we are mainly focused on the case $\alpha=0$. In this limit, while the first family contains only domain wall solutions, within the second family there exist regular black holes and the thermal superpotential can be analytically obtained. The hairy black hole solutions exist only for $\nu>2$ (the special case $\nu=1$ is presented below), otherwise there are naked singularities. While the thermal superpotential has the same qualitative behaviour for any value of $\nu$, there is a drastic change in the behaviour of the dilaton potential, see Fig.\ \subref{subfig:dilPot} and \subref{subfig:thermSuperpot}. This happens because an extremum of the superpotential will automatically be an extremum of the potential, but the converse is not true in general.
\par
The success of AdS-CFT duality is coming from providing a concrete computational `recipe' for relating the bulk (super)gravity/string theory with the dual field theory at the boundary. Particularly, the equivalence of (part of) the bulk and boundary spectra can be explicitly verified. However, there exists a class of  boundary operators with no obvious SUGRA counterpart, namely the
multi-trace gauge invariant operators of $\N=4$ SYM. The existence of these operators in the dual field theory posed a puzzle, since they arise in the operator product expansions of the boundary operators at strong coupling \cite{Arutyunov:2000ku,Arutyunov:2000im} and so they should also have an interpretation in the bulk supergravity framework. Now it is well understood that they can by studied via AdS-CFT by a generalization of the boundary conditions. In this context, the hairy black holes presented in our paper -- which correspond to mixed boundary conditions for the dilaton -- can be interpreted as triple trace deformations in the dual field theory.

To obtain the dual RG flow, it is important to obtain the dynamics of the system as first order flow equations, and this can be done by introducing the (fake) superpotential. The key observation of \cite{deBoer:1999tgo} is that the equations of motion are in fact the Hamilton-Jacobi equations for the dynamical system of gravity and scalars, and the superpotential is nothing else than the classical Hamilton-Jacobi function. However, since we consider hairy black hole solutions, we have obtained the corresponding thermal superpotential and, with the help of the `bulk-boundary dictionary' of \cite{deBoer:1999tgo}, we had constructed the exact RG flow in Section \ref{sec:RGflow}.\par
Since all the details were presented carefully, we would only like to point out some properties of the $c$-function (\ref{cfuncionBH}). The central charge counts the number of massless degrees of freedom in the CFT. The coarse graining of a quantum field theory removes the information about the small scales, so that there is a gradual loss of non-scale invariant degrees of freedom. This is basically the reason behind the existence of a $c$-function that is decreasing monotonically from the UV regime (or large radii in the dual AdS space) to the IR regime (or small radii in the gravity bulk dual) of the QFT. We emphasize that the $c$-function depends only on the conformal factor, not on the metric function, which is consistent with the fact that we deal with the same theory, but at finite temperature. Then, we notice that, when the hairy parameter has the value $\nu=1$, the moduli metric vanishes and we obtain the Schwarzschild-AdS solution for which the flow is trivial.  In this case, we have ${\mathcal{C}(\nu=1)=\mathcal{C}_0\,L^2}$ and so, in principle, the constant $\mathcal{C}_0$ can be computed in this limit.%
\footnote{If we consider the number of parallel branes, $N$, and the embedding in string theory, the number of `unconfined' degrees of freedom for AdS$_4$ is $N^{3/2}\sim \ell^2/G$.}\par
Examples of flows between two conformal fixed points and flows to massive theories can be found in \cite{Distler:1998gb, Girardello:1998pd, Freedman:1999gp}.  Note that on the gravity side, the RG solutions corresponding to the flow to massive theories are generically singular. In our case, the near horizon geometry does not contain an AdS$_2$ spacetime as in the case of zero temperature and so the horizon of planar black hole at finite temperature is not an IR critical point, but, due to the existence of the horizon, it is not singular. Exact charged hairy black holes in extended supergravity, for which a similar analysis at zero temperature is possible, are going to be presented in \cite{nosotros1}. Similar examples, but in a different context, were presented in \cite{Guarino:2017eag,Guarino:2017pkw} and the holographic microstate counting in AdS$_4$ was done in \cite{Azzurli:2017kxo} (see, also, \cite{Zaffaroni:2019dhb} and references therein).

\section*{\normalsize Acknowledgments}
\vspace{-5pt}
Research of AA is supported in part by Fondecyt Grants 1141073 and 1170279. The  research  of  DA  is  supported  by the  Fondecyt Grants  1200986,  1170279,  1171466,  and  2019/13231-7  Programa  de  Cooperacion  Internacional, ANID.
DC is supported by Fondecyt Postdoc Grant 3180185.


\newpage

\hypersetup{linkcolor=blue}
\phantomsection 
\addtocontents{toc}{\protect\addvspace{4.5pt}}
\addcontentsline{toc}{section}{References} 
\bibliographystyle{JHEP}
\bibliography{bibliographyBH} 

\end{document}